\documentclass[prd,showpacs,twocolumn,superscriptaddress,floatfix,nofootinbib,10pt]{revtex4-2}
\usepackage{setspace}
\usepackage[utf8]{inputenc}
\usepackage{float}
\usepackage{amsmath,amssymb,amsfonts,bm}
\usepackage{graphicx}
\usepackage{multirow}
\usepackage[usenames,dvipsnames]{color}
\allowdisplaybreaks
\usepackage[
colorlinks=true,
linkcolor=blue,
breaklinks=true,
urlcolor=blue,
citecolor=blue]{hyperref}

\usepackage{subfigure}
\usepackage{soul}
\usepackage{xcolor}
\usepackage{physics}
\usepackage{booktabs}
\usepackage{ulem}
\usepackage{tikz-feynman}
\usepackage{makecell}
\definecolor{green}{rgb}{0,0.6,0}
\newcommand{\mev}{\textrm{ MeV}}

\begin{document}
	\title{Correlation function and bound state from the $K D_{s0}^*(2317)$ interaction}
	
	\begin{abstract}
In anticipation of the new wave of ALICE experiments on particle–resonance correlation functions, we study the interaction of a kaon with the $D_{s0}^*(2317)$ resonance. Assuming the $D_{s0}^*(2317)$ to be a $DK$ molecular state in isospin $I=0$, we employ the fixed center approximation (FCA) to describe the kaon scattering off the $DK$ cluster, and implement elastic unitarity in the $K D_{s0}^*(2317)$ amplitude via an optical potential and the Lippmann–Schwinger equation. We evaluate the scattering length, effective range, and correlation function, which exhibits a shape characteristic of a strongly attractive interaction. Notably, the amplitude develops a narrow resonant peak about $40\ \mathrm{MeV}$ below the $K D_{s0}^*(2317)$ threshold, signaling a three-body bound state. We discuss the experimental feasibility of observing this state through the invariant mass distribution of $K D_s^+ \pi^0$, and argue that such three-body states, predicted by various theoretical approaches, offer promising targets for future experimental searches, providing valuable insights into the nature of exotic hadronic resonances.
	\end{abstract}

\author{Wen-Hao Jia}
\affiliation{Department of Physics, Guangxi Normal University, Guilin 541004, China}
\affiliation{Guangxi Key Laboratory of Nuclear Physics and Technology,
Guangxi Normal University, Guilin 541004, China}

\author{Hai-Peng Li}
\affiliation{Department of Physics, Guangxi Normal University, Guilin 541004, China}
\affiliation{Guangxi Key Laboratory of Nuclear Physics and Technology,
Guangxi Normal University, Guilin 541004, China}

\author{Wei-Hong Liang}
\affiliation{Department of Physics, Guangxi Normal University, Guilin 541004, China}
\affiliation{Guangxi Key Laboratory of Nuclear Physics and Technology,
Guangxi Normal University, Guilin 541004, China}

\author{Jing Song}
\email[]{Song-Jing@buaa.edu.cn}
\affiliation{Center for Theoretical Physics, School of Physics and Optoelectronic Engineering, Hainan University, Haikou 570228, China}
\affiliation{Department of Physics, Guangxi Normal University, Guilin 541004, China}

\author{ Eulogio Oset}
\email[]{oset@ific.uv.es}
\affiliation{Department of Physics, Guangxi Normal University, Guilin 541004, China}
\affiliation{Departamento de Física Teórica and IFIC, Centro Mixto Universidad de Valencia-CSIC Institutos de Investigación de Paterna, 46071 Valencia, Spain}	

\maketitle

\section{Introduction}\label{sec:Intr}
The ALICE collaboration has started a new wave of experiments investigating the correlation functions stemming from the interaction of stable particles with resonances. 
The first steps in this direction involve the $p f_1(1285)$ system, where the $f_1(1285)$ is detected via the $K \bar K \pi$ decay mode \cite{Privatecomm,ALICE:2024rjz}. 
The aim of the project is to learn about the nature of many resonances which are the subject of permanent debate (see review papers on this issue \cite{Brambilla:2010cs,Chen:2016spr,Dong:2017gaw,Esposito:2016noz,Hosaka:2016pey,Lebed:2016hpi,Olsen:2014qna,Olsen:2017bmm,Oset:2016lyh,Guo:2017jvc}). 
Correlation functions offer a possibility to study the interaction of a particular resonance with many different sources, accumulating novel information that should shed light on the issue of the nature of the resonance investigated. 
Among multiple resonances to be investigated, some of them play a particular role since they can correspond to exotic states which cannot be cast into the standard $q \bar q$ or $3q$ nature for mesons and baryons respectively, or correspond to dynamically generated, molecular-like states that emerge from the interaction of other particles, usually within coupled channels \cite{Guo:2017jvc,Oset:2016lyh}. 
This is the case for the $f_1(1285)$ state, which is considered to be a molecular state originating from the $K \bar K^*$ - $\bar K K^*$ interaction with isospin $I = 0$ \cite{Lutz:2003fm,Roca:2005nm,Garcia-Recio:2010enl,Zhou:2014ila,Geng:2015yta,Lu:2016nlp}. 
A long list of tests supporting this nature is discussed in Ref.~\cite{Jia:2026dpl}.
Theoretical work following this new trend has started to appear, and in Ref.~\cite{Encarnacion:2025lyf} the correlation function for the $p f_1(1285)$ interaction was evaluated, finding a significant deviation from unity which should be contrasted with the coming experimental results \cite{ALICE:2024rjz}.

The work of Ref.~\cite{Encarnacion:2025lyf} used the popular Fixed Center Approximation (FCA) to the Faddeev equations \cite{Kamalov:2000iy,Foldy:1945zz,Brueckner:1953zz,Brueckner:1953zza,Chand:1962ec,Barrett:1999cw,Deloff:1999gc,MartinezTorres:2020hus,Roca:2010tf,Malabarba:2024hlv} in which there is an external particle, the proton, interacting with a cluster of two particles, the $f_1(1285)$ as a composite state of $K \bar K^*$ - $\bar K K^*$, where the cluster is assumed to remain unchanged during the interaction, as is generally assumed when one studies the interaction of particles with nuclei \cite{Ericson:1988gk,Seki:1983sh,Nieves:1993ev,Brown:1975di}. 
The work of Ref.~\cite{Encarnacion:2025lyf} faced a new challenge when it was found that the FCA, normally used to look for bound states below the threshold of the external particle and the cluster, did not fulfill elastic unitarity at the threshold, a condition necessary to determine the effective range in the effective range expansion of the three-body amplitude and the correlation function. 
An ad hoc solution was found in that work by multiplying the amplitude by a factor close to unity, which rendered the approach unitary.
The formal solution to this problem was found in Ref.~\cite{Ikeno:2025bsx} when studying the interaction of a neutron with the $\bar D_{s0}^*(2317)$. 
In that case the $\bar D_{s0}^*(2317)$ was also assumed to be generated by the interaction of $\bar D\bar K$. 

In both cases considered in Refs.~\cite{Encarnacion:2025lyf,Ikeno:2025bsx}, a bound state for the three-body system was also found, providing novel information that calls for an experimental search. 
These predictions are tied to the nature of the $f_1(1285)$ and $\bar D_{s0}^*(2317)$ resonances as dynamically generated from the interaction of more elementary particles, and are quite different from predictions assuming these resonances to be elementary fields.
One clear example is shown in the results of Ref.~\cite{Jia:2026dpl}, where the $K f_1(1285)$ interaction was studied along the lines of Ref.~\cite{Ikeno:2025bsx} and the correlation function was evaluated. 
The interaction was found to be strong enough to produce a correlation function diverting sizeably from unity, and to generate a bound state close to threshold. 
On the other hand, it was found in Ref.~\cite{Yan:2023vbh} that assuming the $f_1(1285)$ as an elementary matter field, the interaction was zero. 

More work has followed along these lines, and in Ref.~\cite{Agatao:2025ckp}, as a complement of the $ n \bar D_{s0}^*(2317)$ interaction studied in Ref.~\cite{Ikeno:2025bsx}, the correlation functions for the $n\,\bar{D}_{s1}(2460)$ and $n\,\bar{D}_{s1}(2536)$ systems were evaluated, with bound states also found. 
A further application of the new unitary techniques to the three-body interaction was presented in Ref.~\cite{Jia:2025obs}, which predicts a super exotic bound state of $K^{*+} D^{*+} K^{*+}$ nature with total spin $J=3$. 

As a complement of the work done in Ref.~\cite{Ikeno:2025bsx}, in the present work we study the interaction of a kaon with the $D_{s0}^*(2317)$ resonance. 
The $D_{s0}^*(2317)$ is also assumed to be generated from the interaction of the $DK$ and $ D_s\eta$ channels \cite{vanBeveren:2003kd,Barnes:2003dj,Chen:2004dy,Kolomeitsev:2003ac,Gamermann:2006nm,Guo:2006rp,Yang:2021tvc,Liu:2022dmm,Liang:2025jkj}, mostly $DK$ in isospin $I=0$, which is also supported by lattice QCD calculations \cite{Mohler:2013rwa,Lang:2014yfa,Bali:2017pdv,Cheung:2020mql,MartinezTorres:2014kpc}. 
This is a new method to study the nature of the $D_{s0}^*(2317)$ since in Ref.~\cite{Ikeno:2025bsx} one is faced with the interaction of the neutron with $\bar K$ and $\bar D$, while here one is considering the interaction of a kaon with $K$ and $D$. These interactions are completely different from the dynamical point of view, so the predictions should be very different and will provide complementary information on the nature of the $D_{s0}^*(2317)$ resonance, which is one of the motivations behind the work. 

\section{Formalism}

We follow closely the work of {{Ref.~\cite{Jia:2026dpl}}}, which contains the developments and improvements of the FCA as implemented in Refs.~\cite{Ikeno:2025bsx,Agatao:2025ckp,Jia:2025obs}.

\begin{figure}[H]
	\begin{center}
		\includegraphics[width=0.48\textwidth]{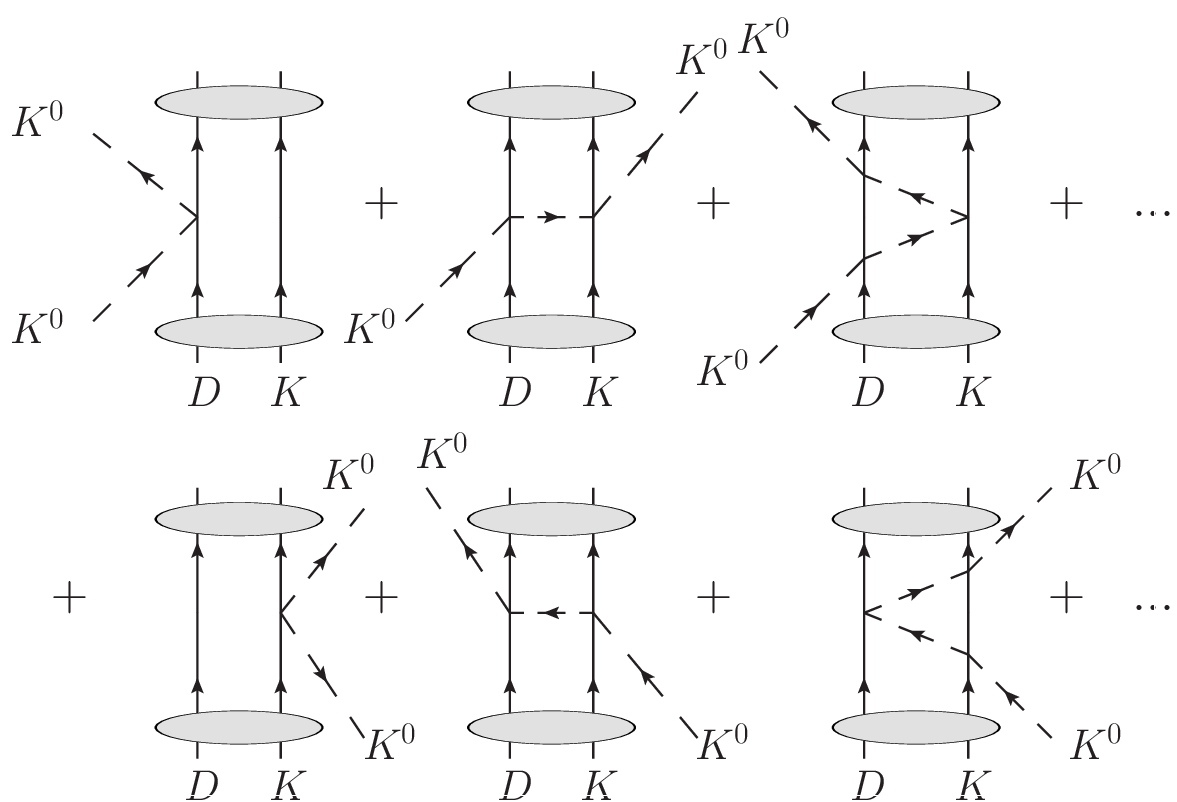}
	\end{center}
	\vspace{-0.5cm}
	\caption{Diagrams entering the FCA approach for $K^0$ interacting with the cluster $DK$.}
	\label{Fig1_1}
\end{figure}

We study the interaction of a kaon with the $D_{s0}^*$(2317) assumed to be a molecular state of $DK$ in isospin $I=0$. The first step is to sum the diagrams involved in the conventional FCA, which are depicted in Fig.~\ref{Fig1_1}.\\

The external $K^0$ can interact with the $D$ and the $K$ of the cluster. The $DK$ interaction is attractive, in fact it produces the $D_{s0}^*(2317)$, while the $KK$ interaction is repulsive, but given the large strength of the $DK$ attraction, one expects that this attraction will overcome the repulsion and lead to a three-body bound state of $DKK$ nature. Let us call $t_1$, $t_2$ the amplitudes for collision of a $K^0$ with the $D$ and $K$ components of the $D_{s0}^*(2317)$ cluster, respectively. When considering the $I=0$ structure of the $DK$ cluster, the amplitudes $t_1$, $t_2$ are written as
\begin{equation}
	\begin{split}
    t_1 &= \frac{3}{4}t_{DK}^{I=1} + \frac{1}{4}t_{DK}^{I=0},\\
    t_2 &= \frac{3}{4}t_{KK}^{I=1} + \frac{1}{4}t_{KK}^{I=0}.
	\end{split}
\end{equation}
For reasons of normalization, by referring the final amplitudes to the interaction of $K^0$ with the cluster as a whole of mass $M_C$, we define the amplitudes
\begin{align}
    \tilde{t}_1=\frac{M_C}{m_D}t_1, ~~~~~ \tilde{t}_2=\frac{M_C}{m_K}t_2.
\end{align}

The arguments of these amplitudes are given by 
\begin{align}
    s_1(DK) &= (p_{K^0}+p_D)^2 = m_K^2 + (\xi m_D)^2 + 2\xi m_D q_0\\\nonumber
    s_2(KK) &= (p_{K^0}+p_K)^2 = m_K^2 + (\xi m_K)^2 + 2\xi m_K q_0,
\end{align}
with $q_0$ the energy of $K^0$ in the rest frame of the cluster
\begin{align}
    q_0 = \frac{s-m_{K^0}^2-M_C^2}{2M_C} ,\qquad \xi = \frac{M_C}{m_D+m_K},
\end{align}
which assume that the binding energy of the $D_{s0}^*$(2317) is shared between the $D$ and $K$ of the cluster proportional to their respective masses.

From Fig.~\ref{Fig1_1} it is useful to separate the terms $\tilde{T}_{ij}$ which includes all diagrams where the $K^0$ interacts first with particle $i$ of the cluster and finishes with particle $j$. One finds
\begin{equation}\label{eq:tij}
	\begin{split}
    &\tilde{T}_{11}=\frac{\tilde{t}_1}{1-\tilde{t}_1\tilde{t}_2G_0^2},~~~~ \tilde{T}_{22}=\frac{\tilde{t}_2}{1-\tilde{t}_1\tilde{t}_2G_0^2},\\
    &\tilde{T}_{12}=\tilde{T}_{21}=\frac{\tilde{t}_1\tilde{t}_2G_0}{1-\tilde{t}_1\tilde{t}_2G_0^2},
	\end{split}
\end{equation}
where $G_0$ is the $K^0$ propagator from one particle to the other in the cluster, given by
\begin{equation}\label{eq:G0}
\begin{aligned}
    G_0(\sqrt{s}) &= \int\frac{d^3 q}{(2\pi)^3}\frac{1}{2\omega_{K}(\boldsymbol{q})}
    \frac{1}{2\omega_C(\boldsymbol{q})}
    \times\frac{F_C(\boldsymbol{q})}{
    \sqrt{s}-\omega_{K}(\boldsymbol{q})-\omega_C(\boldsymbol{q})+i\epsilon}\\
    &\times \Theta(q_{\text{max}}^{(1)} - q_1^*)\Theta(q_{\text{max}}^{(2)} - q_2^*),
\end{aligned}
\end{equation}
where $\omega_K(q)=\sqrt{\bm{q}^2+m_K^2}$, $\omega_C(q)=\sqrt{\bm{q}^2+M_C^2}$.
The presence of the $\Theta (~)$ functions in Eq.~\eqref{eq:G0} stems from the $\bm{q},\bm{q}'$ dependence of the amplitudes~\cite{Gamermann:2009uq}
\begin{align}
    t(\bm{q},\bm{q}') = t ~\Theta(q_{\text{max}}-|\bm{q}|)\Theta(q_{\text{max}}-|\bm{q}'|),
\end{align}
resulting from a separable potential $V(\bm{q},\bm{q}')=V~\Theta(q_{\text{max}}-|\bm{q}|)\Theta(q_{\text{max}}-|\bm{q}'|)$. In this case, the cutoff $q_{\text{max}}$ regularizes the meson-meson loop functions entering the evaluation of the Bethe-Salpeter equation (see Appendix) and leads to a formalism equivalent to the one used in the chiral unitary approach with the on-shell factorization~\cite{Oller:2000fj}. In Eq.~\eqref{eq:G0}, $q_{\text{max}}^{(1)}$ refers to the $DK$ amplitude and $q_{\text{max}}^{(2)}$ refers to the $KK$ amplitude. Furthermore, $q_1^*$ and $q_2^*$ refer to the momenta in the rest frame of the propagating $K^0$ and the respective particle 1, 2 of the cluster.  A sensible prescription is taken for the internal motion of the particles in the cluster~\cite{Jia:2025obs}, following {{\color{black}{Refs.}~\cite{Oller:2000fj,Carrasco:1991we,Boffi:1991nh}}} and the external momentum is assumed to be zero, and then we find,
\begin{align}
    \bm{q}_i^*=\bm{q}\left(1- \frac{1}{2}\frac{m_{K^0}}{m_{K^0}+m_i}\right).
\end{align}

In addition, $F_C(\bm q)$ is the form factor of the cluster, which stems from the wave function of the cluster in momentum space~\cite{Gamermann:2009uq}
\begin{align}
    \psi(p)=g\frac{\Theta(q_{\text{max}}-|\bm{p}|)}{M_C-\omega_D(\bm{q})-\omega_K(\bm{q})},
\end{align}
which leads to 
\begin{equation}\label{eq:FF}
   \nonumber F_C(\boldsymbol{q}) = \frac{F(\boldsymbol{q})}{\mathcal{N}},
    \end{equation}
\begin{align}
    \nonumber F(\boldsymbol{q})=\int\limits_{\substack{|\boldsymbol{p}| < q_{\text{max}}, \\ |\boldsymbol{p - q}| < q_{\text{max}}}} \!\! &\frac{d^3p}{(2\pi)^3}\frac{1}{M_C - \omega_{D}(\boldsymbol{p})-\omega_{K} (\boldsymbol{p})} \\\nonumber
    & \times \frac{1}{M_C - \omega_{D}(\boldsymbol{p-q})-\omega_{K}(\boldsymbol{p-q})},
\end{align}
\begin{equation}
    \mathcal{N}=F(0)=\int\limits_{|\boldsymbol{p}|<q_{\text{max}}}\frac{d^3p}{(2\pi)^3}\left( \frac{1}{M_C - \omega_{D}(\boldsymbol{p})-\omega_{K}(\boldsymbol{p})}\right)^2.
\end{equation}

\begin{figure}[H]
	\begin{center}
		\includegraphics[width=0.48\textwidth]{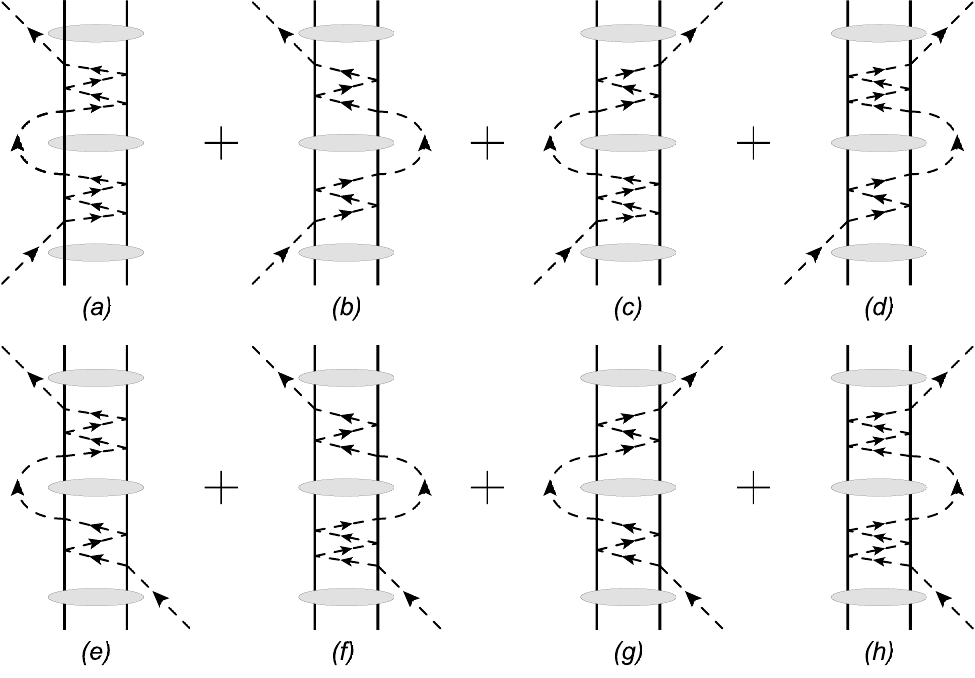}
	\end{center}
	\vspace{-0.5cm}
	\caption{Diagrams considering the elastic propagation of the $K^0$ and the cluster $D_{s0}^*(2317)$.}
	\label{Fig2_1}
\end{figure}

For later purposes, it is  interesting  to put $\tilde{T}_{ij}$ of Eq.~\eqref{eq:tij} in form of a matrix, 
\begin{align}
\tilde{T}&=
\begin{pmatrix}	
\tilde{T}_{11} & \tilde{T}_{12} \\	
\tilde{T}_{21} & \tilde{T}_{22}
\end{pmatrix}.
\end{align}
Next we introduce the unitarization in the $K^0D^*_{s0}(2317)$ system, with the  $D^*_{s0}(2317)$ taken as a whole, which proceeds by incorporating the diagrams shown in Fig.~\ref{Fig2_1}.
The need for the diagrams of Fig.~\ref{Fig2_1} recalls what happens in the interaction of an external particle with a nucleus, where the $t_\rho$ ($\rho$ the density of the nucleus), (which would be equivalent to $t_1+t_2$ in our formalism in the sum of the diagrams of Fig.~\ref{Fig1_1}) provides the optical potential, which must be used in the Lippmann-Schwinger equation to obtain the particle-nucleus scattering amplitude. 

The sum of diagrams in Fig.~\ref{Fig2_1}, including extra iterations in the $K^0D_{s0}^*(2317)$ propagation can be represented  in terms of the equation
\begin{align}
    \tilde{T}'=[1-\tilde{T}G_C]^{-1}\tilde{T},
\end{align}
with 
\begin{align}
    G_C=\begin{pmatrix}
          G^{(1)}_{C} & 0 \\ 0 & G^{(2)}_{C} 
      \end{pmatrix},
\end{align}
where 
\begin{equation}\label{eq:Gcs}
	\begin{aligned}
    G^{(i)}_{C}(\sqrt{s})&=\int\frac{d^3 q}{(2\pi)^3}\frac{1}{2\omega_K(\boldsymbol{q})}\frac{1}{2\omega_C(\boldsymbol{q})}\\ &\times\frac{[F^{(i)}_C(\boldsymbol{q})]^2}{\sqrt{s}-\omega_K(\boldsymbol{q})-\omega_C(\boldsymbol{q})+i\epsilon}\Theta(q_{\text{max}}^{(i)}-q_i^*),
\end{aligned}
\end{equation}
with {\cite{Yamagata-Sekihara:2010kpd}}
\begin{equation}
	\begin{split}
    F^{(1)}_C(\boldsymbol{q}) &= F_C\left(\frac{m_{K}}{m_{{D}}+m_{{K}}}\boldsymbol{q}\right),\\
    F^{(2)}_C(\boldsymbol{q}) &= F_C\left(\frac{m_{{D}}}{m_{{D}}+m_{{K}}}\boldsymbol{q}\right). 
	\end{split}
\end{equation}
The final amplitude is then given by \cite{Agatao:2025ckp}
\begin{equation}\label{eq:Alberto_T}
\begin{aligned}
    T^{\text{tot}}(\sqrt{s})   &=\sum_{i,j} \tilde{T}_{ij}^{~'} \\
    & =  \frac{\tilde{t}_1+\tilde{t}_2+(2G_0-G_C^{(1)}-G_C^{(2)})\tilde{t}_1\tilde{t}_2}{1-G_C^{(1)}\tilde{t}_1-G_C^{(2)}\tilde{t}_2-(G_0^2-G_C^{(1)}G_C^{(2)})\tilde{t}_1\tilde{t}_2}.
\end{aligned}
\end{equation}

\subsection{Scattering length and effective range}
Once we have the total amplitude $T^{\text{tot}}(\sqrt{s})$, one can obtain the $K^0D_{s0}^*(2317)$ scattering length and effective range. Taking into account the normalization of our amplitude compared to the usual one employed in Quantum Mechanics, we have
\begin{equation}\label{eq:ar}
\begin{aligned}
   - 8\pi\sqrt{s} (T^{\text{tot}})^{-1}&= (f^{\text{QM}})^{-1} \\
   &\simeq-\frac{1}{a}+\frac{1}{2}r_0 q_{\text{cm}}^2 - i q_{\text{cm}},
\end{aligned}
\end{equation}
where $q_{\text{cm}}$ is the $K^0$ momentum in the $K^0D_{s0}^*(2317)$ rest frame. The term $-i q_{\text{cm}}$ in Eq.~\eqref{eq:ar} implies the elastic unitarity of the $K^0D_{s0}^*(2317)$ amplitude, and it was shown analytically in Ref.~\cite{Agatao:2025ckp} that the amplitude of Eq.~\eqref{eq:Alberto_T} fulfills exactly this unitarity. Then one obtain 
\begin{align}
    a&= \left.{\frac{T^{\text{tot}}}{8\pi\sqrt{s}}} \right|_{\text{th}} , \\
    r_0&=\frac{1}{\mu}\left[\frac{\partial}{\partial\sqrt{s}}\left(-{8\pi\sqrt{s}}(T^{\text{tot}})^{-1} {+ iq_{\text{cm}}}\right)\right]_\text{th},
\end{align}
with $\mu$ the $K^0D_{s0}^*(2317)$ reduced mass.
\subsection{Correlation function}
Following Ref.~\cite{Jia:2026dpl}  we write the correlation function
\begin{equation}
\begin{aligned}
    C_{DK_{s0}^*}(p) = 1 + 4\pi &\int_0^{\infty}  dr\, r^2 S_{12}(r) \\
  &\times\left\{ \left| j_0(pr) + TG \right|^2
- j_0^2(pr) \right\},
\end{aligned}
\end{equation}
where 
\begin{align}
    TG =& (\tilde{T}^{~'}_{11} + \tilde{T}^{~'}_{21}) G_1(\sqrt{s},r)   
    + (\tilde{T}^{~'}_{12} +  \tilde{T}^{~'}_{22}) G_2(\sqrt{s},r)  
\end{align}
with $S_{12}(r)$ the source function, given by 
\begin{align}
       S_{12}(r) = \frac{1}{(4\pi R^2)^{3/2}}\exp(-r^2/4R^2),
\end{align}
and $R$ the radius of the source, and 
\begin{equation}
\begin{aligned}
     G_{i}(\sqrt{s},r)=\int&\frac{d^3 q}{(2\pi)^3}\frac{1}{2\omega_K(\boldsymbol{q})}\frac{1}{2\omega_C(\boldsymbol{q})}\Theta(q_{\text{max}}^{(i)}-q_i^*)\\ &\times\frac{j_0(qr)F^{(i)}_C(\boldsymbol{q})}{\sqrt{s}-\omega_K(\boldsymbol{q})-\omega_C(\boldsymbol{q})+i\epsilon}. 
\end{aligned}
\end{equation}

\section{Results}\label{sec:res}
We first show the results obtained for the scattering length and the effective range. We find
\begin{align}
	a &= (0.517 - 0.00016 \,i )\,\mathrm{fm}, 
	\label{eq:24} \\
	r_0 &= (-0.511 + 0.102 \, i)\,\mathrm{fm}. 
	\label{eq:25}
\end{align}
The small imaginary part of $a$ reflects the small imaginary part of the $DK$ amplitude arising from the small width of the $D_{s0}^*(2317)$.
Both $a$ and $r_0$ have magnitudes of about half of those of the $pf_1(1285)$ scattering, but the sign of $r_0$ here is opposite to that found in $pf_1(1285)$.
They are, however, much closer to those obtained for the $Kf_1(1285)$ interaction in Ref.~\cite{Jia:2026dpl}.

\begin{figure}[tb]
	\centering
	\subfigure[]{
		\includegraphics[width=0.48\textwidth]{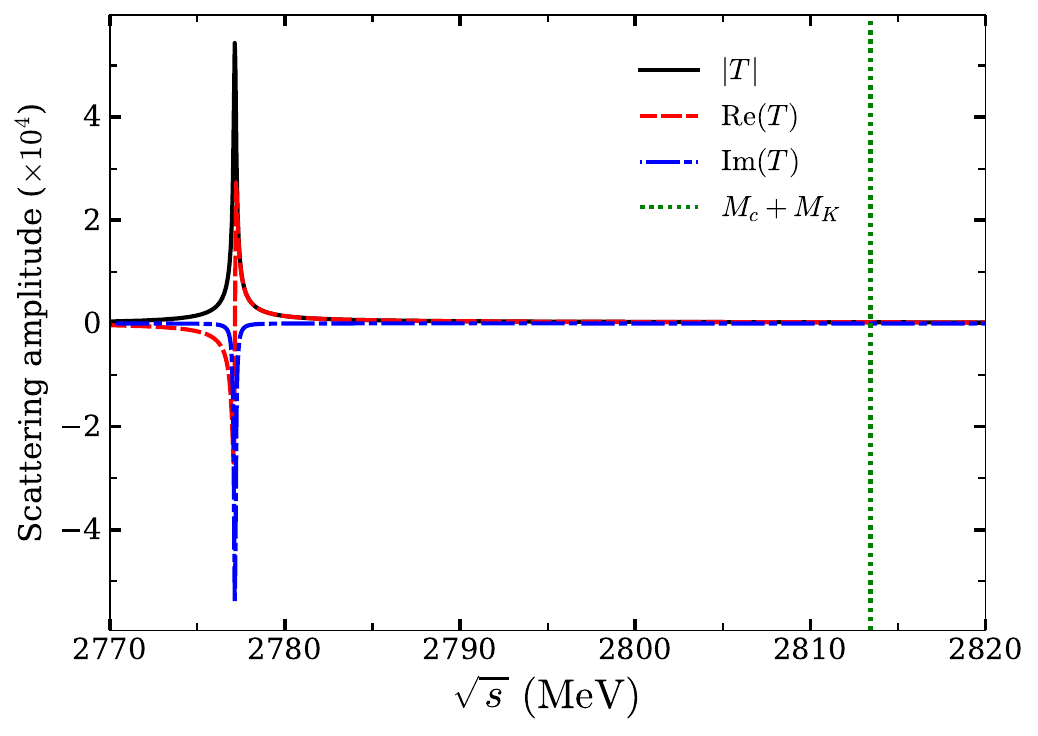} 
		\label{fig3a} 
	}\\
	\vspace{-0.2cm} 
	\subfigure[]{
		\includegraphics[width=0.48\textwidth]{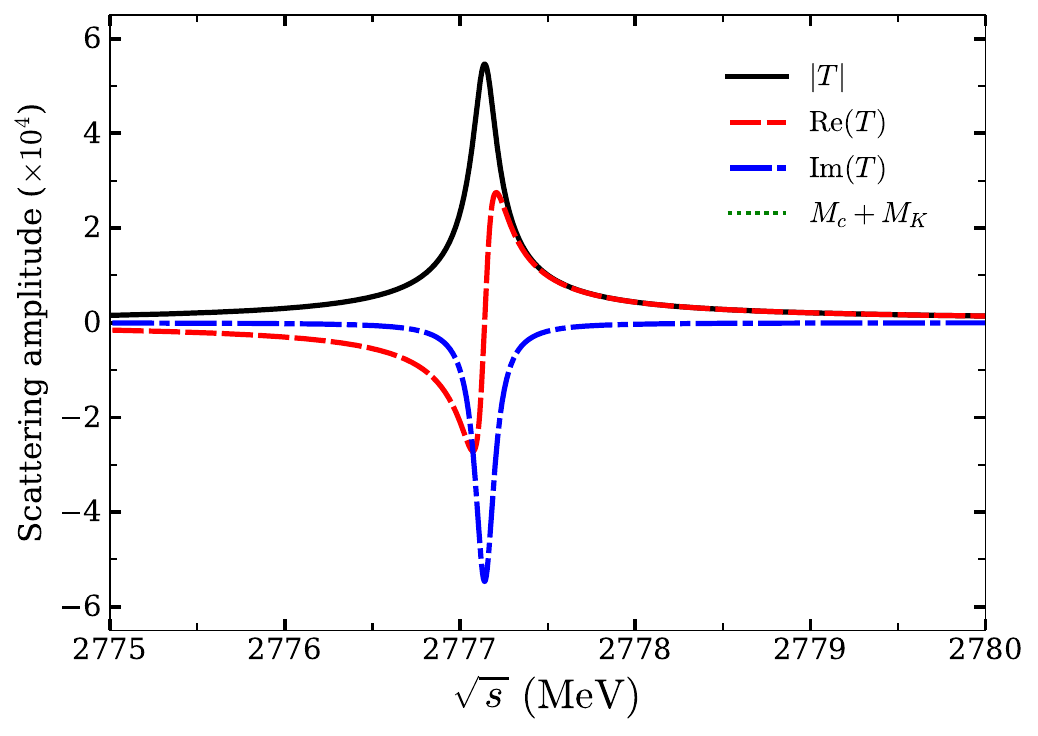}
		\label{fig3b}
	}
	\vspace{-0.4cm} 
	\caption{Results for $T^{\mathrm{tot}}\left(\sqrt{s}\right)$ as a function of $\sqrt{s}$: (a) relative to the $DK_{s0}^*(2317)$ threshold, (b) blow up in a narrow energy window.}
	\label{fig3}
\end{figure}

Next we show the results for the $T^{\mathrm{tot}}$ matrix as a function of the center of mass energy of the $DK_{s0}^*(2317)$ system, $\sqrt{s}$, in Fig.~\ref{fig3}.
We find a structure typical of a narrow resonance around $40\mev$ below the $DK_{s0}^*(2317)$ threshold.
The modulus of the imaginary part has a peak around $2777\mev$ and the real part goes through zero around that energy.
The absolute value of $T^{\mathrm{tot}}$ shows a neat and narrow peak.
From $\left| \Im\, T\right|$ at half height of its peak we find the width of the state to be around $200\,\mathrm{keV}$.
The source of the state found stems from the $DK$ attraction.
We have checked the effect of the $KK$ repulsion. If we remove $t_{KK}$ from the approach, the state appears at $2771\mev$.
The effect of the repulsion has been to move the position of the state $6\mev$ closer to the $DK_{s0}^*(2317)$ threshold.

One might be surprised that the binding energies obtained and the width are bigger than those in the $DK$ system itself.
This is an effect produced by expressing the results in terms of the invariant mass distribution of particle-cluster, $\sqrt{s}$, rather than particle-particle, $\sqrt{s_1}$, an effect well known in particle-nucleus collisions \cite{Montesinos:2024eoy}.

\begin{figure}[t]
	\begin{center}
			\includegraphics[width=0.48\textwidth]{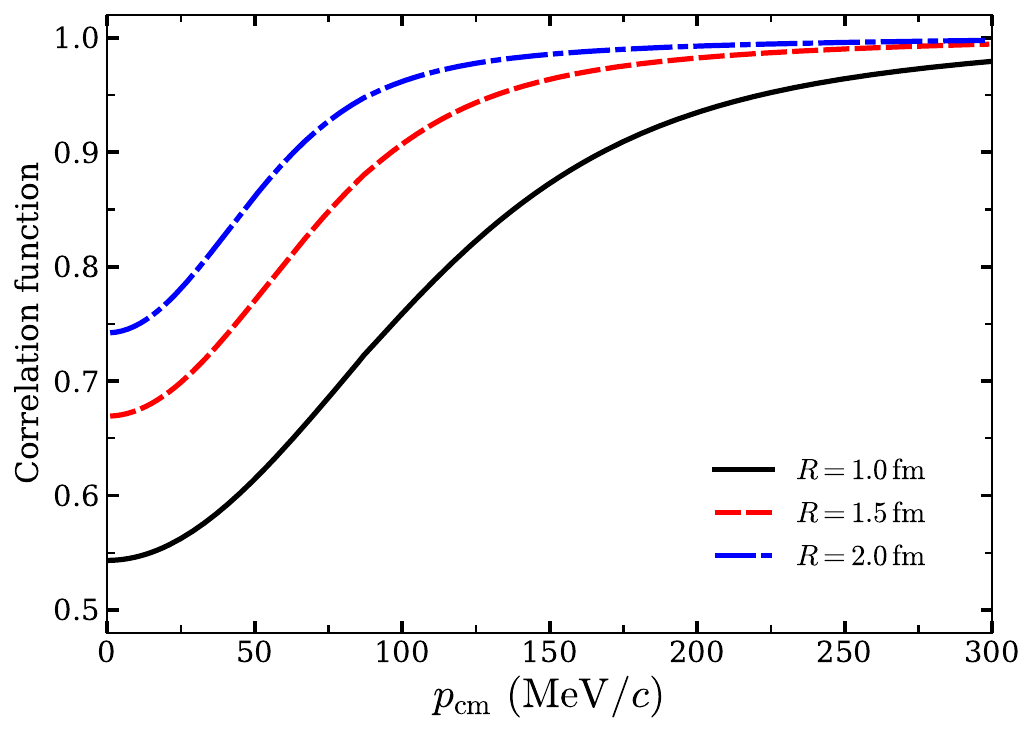}
		\end{center}
	\vspace{-0.5cm}
	\caption{Correlation function for the $K D_{s0}^*(2317)$ system.}
	\label{fig4}
\end{figure}

Since the $D_{s0}^*(2317)$ decays into $D_s^+ \pi ^0$, the peak that we obtained should be searched in the invariant mass distributions of $K^0 D_s^+ \pi ^0$.
A detailed discussion on how such invariant mass distributions can be searched in present facilities by the LHCb or ALICE collaborations is given at the end of section III in Ref.~\cite{Jia:2025obs}.
Next we turn to the correlation function of the $DK_{s0}^*(2317)$ system. The results are shown in Fig.~\ref{fig4}.

We observe a pattern very similar to the one found for the $pf_1(1285)$ in Ref.~\cite{Encarnacion:2025lyf}, and the $Kf_1(1285)$ in Ref.~\cite{Jia:2026dpl}, all of which produce a bound state with similar binding energies.
In the qualitative picture for correlation functions discussed in Refs.~\cite{Liu:2023uly,Liu:2024uxn}, it would correspond to the case of a strongly attractive potential.
The shapes of the correlation functions in Fig.~\ref{fig4} are also similar to those obtained for the $D^0K^+$ and $D^+K^0$ correlation functions in Ref.~\cite{Ikeno:2023ojl}.

As in the case of the $pf_1(1285)$ where the $f_1(1285)$ is identified by a decay channel, the $K\bar K \pi$ , in the measurement of the $K^0D_{s0}^*(2317)$ correlation function, the $D_{s0}^*(2317)$ would be identified by its $D_s^+ \pi^0$ decay, by looking for the peak of the invariant mass distribution of $D_s^+ \pi^0$ to identify the peak corresponding to the $D_{s0}^*(2317)$ state, while a $K^0$ is identified as coming from the same event.
With the same  experimental machinery, one could look at the invariant mass distribution of $K^0D_s^0 \pi^0$ below threshold to identify the state predicted.
We think that there is much to search, and hopefully find, by making a thorough measurement of three-body invariant mass distribution in this and other related cases.

\section{Conclusions}
Anticipating results in the new wave of ALICE experiments on particle-resonance correlation functions, we have carried out the evaluation of observables related to the interaction of a kaon with the $D_{s0}^*(2317)$ resonance. 
For this purpose we have assumed that the $D_{s0}^*(2317)$ is a molecular state of $DK$ in isospin $I=0$. 
Then we have used a version of the fixed center approximation, in which the kaon interacts with the $DK$ cluster without destroying it, and made a mapping onto the conventional interaction of particles with nuclei, which constructs an optical potential that is later used within the Lippmann-Schwinger equation to obtain the particle-nucleus scattering matrix. 
The FCA provides the ``optical potential", and the propagation of the elastic channel $K D_{s0}^*(2317)$ is implemented, producing elastic unitarity in the $K D_{s0}^*(2317)$ scattering amplitude, which is essential to evaluate scattering observables at low energies.

We have then evaluated the scattering length, the effective range and the correlation function, which has a shape corresponding qualitatively to a strongly attractive potential. 

As a side product, we observe that the amplitude develops a very narrow resonant shape below the $K D_{s0}^*(2317)$ threshold, corresponding to a three-body state bound by about $40\, {\rm MeV}$. 
We also discuss that while the measurement of this correlation function requires the observation of a kaon and the $K D_{s0}^*(2317)$ from the same event, the $D_{s0}^*(2317)$ being detected via a peak in the $D^+_{s} \pi^0$ mass distribution, the same experimental framework could look into the invariant mass distribution of three bodies, $K D^+_{s} \pi^0$ below threshold, looking for the predicted state. 
We anticipate that this search, and similar ones of predicted three-body states, are bound to find many three-body states which are predicted by different theoretical approaches (see review on this issue in Ref.~\cite{MartinezTorres:2020hus}). 

\section*{Acknowledgments}
 We would like to thank Prof. Natsumi Ikeno for useful information.
This work is partly supported by the National Natural Science Foundation of China (NSFC) under Grants No. 12575081 and No. 12365019,
and by the Natural Science Foundation of Guangxi province under Grant No. 2023JJA110076,
and by the Central Government Guidance Funds for Local Scientific and Technological Development, China (No. Guike ZY22096024). 
This work is also partly supported by the National Key R{\&}D Program of China (Grant No. 2024YFE0105200).
J. Song acknowledges support from the
National Natural Science Foundation of China (NSFC)
under Grants No. 12405089 and No. 12247108, and by the China Postdoctoral Science Foundation under Grant No. 2022M720360
and No. 2022M720359. J. Song would also like to thank the support from the Hainan Provincial Excellent Talent Team under the “Four Talents” Gathering Program of Hainan Province.
This work is also partly supported by the Spanish Ministerio de Economia y Competitividad (MINECO) and European FEDER funds under Contracts No. FIS2017-84038-C2-1-PB, PID2020-112777GB-I00, and by Generalitat Valenciana under contract PROMETEO/2020/023. 
This project has received funding from the European Union Horizon 2020 research and innovation program under the program H2020-INFRAIA-2018-1, grant agreement No. 824093 of the STRONG-2020 project.

\appendix
\section{$t_1,\,t_2$ amplitudes}

\subsection{The $DK$ amplitudes}

For $t_1$ we need the $DK$ amplitude in $I=0,\,1$.
We take advantage of the recent work done in Ref.~\cite{Su:2025aiz} where the interaction of the coupled channels, $D^0K^+,\,D^+K^0,D_s^+ \eta$ and $D_s^+\pi^0$ are considered, which allows one to obtain the strong decay width of the $D_{s0}^*(2317)$ into the isospin forbidden $D_s^+\pi^0$ channel.
By using the local hidden gauge approach \cite{Bando:1984ej,Bando:1987br,Meissner:1987ge,Nagahiro:2008cv} with the exchange of vector mesons, the transition potentials $V_{ij}$ between these coupled channels are evaluated in Ref.~\cite{Su:2025aiz}, and then the scattering matrix is evaluated with the Bethe-Salpeter equation in matrix form
\begin{equation}\label{eq:A1}
	T=\left[1-VG\right]^{-1}V,
	\end{equation}
with $G$ the diagonal meson-meson loop function, $G=\mathrm{diag}\left[ G_i \right]$, with
\begin{equation}\label{eq:A2}
	\begin{aligned}[b]
		G_i(s) = \int_{|{\vec q\,}| < q_{\rm max}} \,& \dfrac{\dd^3 q}{(2\pi)^3}\,\dfrac{\omega^{(i)}_1(\vec q\,) + \omega^{(i)}_2(\vec q\,)}{2 \,\omega^{(i)}_1(\vec q\, ) \, \omega^{(i)}_2(\vec q\,)} \\[2mm]
		&\times\dfrac{1}{s-[\omega^{(i)}_1(\vec q\,) + \omega^{(i)}_2(\vec q\,)]^2+i\epsilon},
	\end{aligned}
	\end{equation}
with $\omega^{(i)}_j(\vec q\,) = \sqrt{{\vec{q}}^{\,2}+{m^{(i)}_{j}}^{2}}$ for $j=1,\,2$, the two mesons of channel $i$.
A pole is obtained for a mass of $M_R=2317\mev$ and a width of around $100\,\mathrm{keV}$, and the couplings $g_i$ ($i=1$ to $5$) are defined such that the amplitude at the pole is given by
\begin{equation}\label{eq:A3}
	\begin{aligned}[b]
	T_{ij}&=\frac{g_i\,g_j}{s-M_{R}^2+i M_{R}\Gamma_{R}},\\[2mm]
	M_R&=2317\mev,~~~~\Gamma_R \simeq 100\,\mathrm{keV}.
	\end{aligned}
	\end{equation}
The couplings are given in Ref.~\cite{Su:2025aiz} and from them we can get the $I=0$ scattering matrix, considering the intrinsic isospin phase convention that we use, with the isospin doublets $\left(D^+,-D^0\right),\,\left(K^+,K^0\right)$.
The $DK$ isospin states are then
\begin{equation}\label{eq:A4}
	\begin{aligned}
		\left| DK,\,I=0\right>&=\frac{1}{\sqrt{2}}\left(D^+K^0 + D^0K^+\right),\\[1mm]
		\left| DK,\,I=1,\,I_3=0\right>&=\frac{1}{\sqrt{2}}\left(D^+K^0 - D^0K^+\right).
	\end{aligned}
	\end{equation}
Then
\begin{align}\label{eq:A5}
		t^{I=0}
		& =\frac{1}{2}\left(
		t_{D^+K^0,\,D^+K^0}+2\,t_{D^+K^0,\,D^0K^+}+t_{D^0K^+,\,D^0K^+}
		\right)\nonumber\\[1mm]
		& =\frac{1}{2}\frac{1}{
			s-M_{R}^2+ i M_{R}\Gamma_{R}
		}\nonumber\\[1.5mm]
		&\quad \times \left(
		g_{D^+K^0}^2+2\,g_{D^+K^0}\,g_{D^0K^+}+g_{D^0K^+}^2
		\right),
\end{align}
and from Ref.~\cite{Su:2025aiz}
\begin{equation}\label{eq:A6}
	\begin{split}
		g_{D^+K^0}&=\left(
			8129.49+i\,75.70
		\right)\mev,\\[1mm]
		g_{D^0K^+}&=\left(
		8252.26-i\,69.15
		\right)\mev.
	\end{split}
	\end{equation}
We cannot use these couplings to calculate the $I=1$ amplitude since they refer to the $I=0$ state.
Instead we can use the results of Ref.~\cite{Su:2025aiz} for the potentials in the coupled channels, and taking channel 1 as $D^0K^+$ and channel 2 as $D^+K^0$, and we find
\begin{equation}\label{eq:A7}
	V^{I=1}=\frac{1}{2}\left(
		V_{11}+V_{22}-2\,V_{12}
	\right)=0,
	\end{equation}
where the zero results by taking $m_{\omega}=m_{\rho}$. 
Thus
\begin{equation}\label{eq:A8}
	t_{DK}^{I=1}=0.
	\end{equation}

\subsection{The $KK$ amplitudes}
These amplitudes are evaluated in Appendix A of Ref.~\cite{Jia:2026dpl} with the result
\begin{equation}\label{eq:A9}
	\begin{aligned}[b]
		V_{KK}^{I=1}=\frac{4\,g^2}{M_V^2}\left(
		\frac{3}{2}s-2\,M_K^2
		\right),\\
		g=\frac{M_V}{2f},~~~f=93\mev,~~~M_V=800\mev,
	\end{aligned}
	\end{equation}
and
\begin{equation}\label{eq:A10}
		t_{KK}^{I=1}=\dfrac{V_{KK}^{I=1}}{
			1-\frac{1}{2}\,V_{KK}^{I=1}\,G
		},
	\end{equation}
with the factor $\frac{1}{2}$ in $VG$ to account for the identity of the two kaons, and $G$ given again in the cutoff form of Eq.~\eqref{eq:A2} with $q_{\mathrm{max}}=650\mev$.
Similarly the isospin $I=0$ potential $V_{KK}^{I=0}$ is found null and thus
\begin{equation}\label{eq:A11}
	t_{KK}^{I=0}=0.
	\end{equation}

\bibliography{refsn}

\begin{thebibliography}{66}%
\makeatletter
\providecommand \@ifxundefined [1]{%
 \@ifx{#1\undefined}
}%
\providecommand \@ifnum [1]{%
 \ifnum #1\expandafter \@firstoftwo
 \else \expandafter \@secondoftwo
 \fi
}%
\providecommand \@ifx [1]{%
 \ifx #1\expandafter \@firstoftwo
 \else \expandafter \@secondoftwo
 \fi
}%
\providecommand \natexlab [1]{#1}%
\providecommand \enquote  [1]{``#1''}%
\providecommand \bibnamefont  [1]{#1}%
\providecommand \bibfnamefont [1]{#1}%
\providecommand \citenamefont [1]{#1}%
\providecommand \href@noop [0]{\@secondoftwo}%
\providecommand \href [0]{\begingroup \@sanitize@url \@href}%
\providecommand \@href[1]{\@@startlink{#1}\@@href}%
\providecommand \@@href[1]{\endgroup#1\@@endlink}%
\providecommand \@sanitize@url [0]{\catcode `\\12\catcode `\$12\catcode `\&12\catcode `\#12\catcode `\^12\catcode `\_12\catcode `\%12\relax}%
\providecommand \@@startlink[1]{}%
\providecommand \@@endlink[0]{}%
\providecommand \url  [0]{\begingroup\@sanitize@url \@url }%
\providecommand \@url [1]{\endgroup\@href {#1}{\urlprefix }}%
\providecommand \urlprefix  [0]{URL }%
\providecommand \Eprint [0]{\href }%
\providecommand \doibase [0]{https://doi.org/}%
\providecommand \selectlanguage [0]{\@gobble}%
\providecommand \bibinfo  [0]{\@secondoftwo}%
\providecommand \bibfield  [0]{\@secondoftwo}%
\providecommand \translation [1]{[#1]}%
\providecommand \BibitemOpen [0]{}%
\providecommand \bibitemStop [0]{}%
\providecommand \bibitemNoStop [0]{.\EOS\space}%
\providecommand \EOS [0]{\spacefactor3000\relax}%
\providecommand \BibitemShut  [1]{\csname bibitem#1\endcsname}%
\let\auto@bib@innerbib\@empty
\bibitem [{\citenamefont {\v{S}erk\v{s}nyt\.{e}}\ \emph {et~al.}()\citenamefont {\v{S}erk\v{s}nyt\.{e}}, \citenamefont {Kundu},\ and\ \citenamefont {Korwieser}}]{Privatecomm}%
  \BibitemOpen
  \bibfield  {author} {\bibinfo {author} {\bibfnamefont {L.}~\bibnamefont {\v{S}erk\v{s}nyt\.{e}}}, \bibinfo {author} {\bibfnamefont {S.}~\bibnamefont {Kundu}},\ and\ \bibinfo {author} {\bibfnamefont {M.}~\bibnamefont {Korwieser}},\ }\href@noop {} {}\bibinfo {howpublished} {(Private communication)}\BibitemShut {NoStop}%
\bibitem [{\citenamefont {Acharya}\ \emph {et~al.}(2025)\citenamefont {Acharya} \emph {et~al.}}]{ALICE:2024rjz}%
  \BibitemOpen
  \bibfield  {author} {\bibinfo {author} {\bibfnamefont {S.}~\bibnamefont {Acharya}} \emph {et~al.} (\bibinfo {collaboration} {ALICE}),\ }\bibfield  {title} {\bibinfo {title} {{Measurement of $f_1(1285)$ production in $pp$ collisions at $\sqrt{s} = 13$ TeV}},\ }\href {https://doi.org/10.1016/j.physletb.2025.139562} {\bibfield  {journal} {\bibinfo  {journal} {Phys. Lett. B}\ }\textbf {\bibinfo {volume} {866}},\ \bibinfo {pages} {139562} (\bibinfo {year} {2025})},\ \Eprint {https://arxiv.org/abs/2409.11936} {arXiv:2409.11936 [hep-ex]} \BibitemShut {NoStop}%
\bibitem [{\citenamefont {Brambilla}\ \emph {et~al.}(2011)\citenamefont {Brambilla} \emph {et~al.}}]{Brambilla:2010cs}%
  \BibitemOpen
  \bibfield  {author} {\bibinfo {author} {\bibfnamefont {N.}~\bibnamefont {Brambilla}} \emph {et~al.},\ }\bibfield  {title} {\bibinfo {title} {{Heavy Quarkonium: Progress, Puzzles, and Opportunities}},\ }\href {https://doi.org/10.1140/epjc/s10052-010-1534-9} {\bibfield  {journal} {\bibinfo  {journal} {Eur. Phys. J. C}\ }\textbf {\bibinfo {volume} {71}},\ \bibinfo {pages} {1534} (\bibinfo {year} {2011})},\ \Eprint {https://arxiv.org/abs/1010.5827} {arXiv:1010.5827 [hep-ph]} \BibitemShut {NoStop}%
\bibitem [{\citenamefont {Chen}\ \emph {et~al.}(2017)\citenamefont {Chen}, \citenamefont {Chen}, \citenamefont {Liu}, \citenamefont {Liu},\ and\ \citenamefont {Zhu}}]{Chen:2016spr}%
  \BibitemOpen
  \bibfield  {author} {\bibinfo {author} {\bibfnamefont {H.-X.}\ \bibnamefont {Chen}}, \bibinfo {author} {\bibfnamefont {W.}~\bibnamefont {Chen}}, \bibinfo {author} {\bibfnamefont {X.}~\bibnamefont {Liu}}, \bibinfo {author} {\bibfnamefont {Y.-R.}\ \bibnamefont {Liu}},\ and\ \bibinfo {author} {\bibfnamefont {S.-L.}\ \bibnamefont {Zhu}},\ }\bibfield  {title} {\bibinfo {title} {{A review of the open charm and open bottom systems}},\ }\href {https://doi.org/10.1088/1361-6633/aa6420} {\bibfield  {journal} {\bibinfo  {journal} {Rept. Prog. Phys.}\ }\textbf {\bibinfo {volume} {80}},\ \bibinfo {pages} {076201} (\bibinfo {year} {2017})},\ \Eprint {https://arxiv.org/abs/1609.08928} {arXiv:1609.08928 [hep-ph]} \BibitemShut {NoStop}%
\bibitem [{\citenamefont {Dong}\ \emph {et~al.}(2017)\citenamefont {Dong}, \citenamefont {Faessler},\ and\ \citenamefont {Lyubovitskij}}]{Dong:2017gaw}%
  \BibitemOpen
  \bibfield  {author} {\bibinfo {author} {\bibfnamefont {Y.}~\bibnamefont {Dong}}, \bibinfo {author} {\bibfnamefont {A.}~\bibnamefont {Faessler}},\ and\ \bibinfo {author} {\bibfnamefont {V.~E.}\ \bibnamefont {Lyubovitskij}},\ }\bibfield  {title} {\bibinfo {title} {{Description of heavy exotic resonances as molecular states using phenomenological Lagrangians}},\ }\href {https://doi.org/10.1016/j.ppnp.2017.01.002} {\bibfield  {journal} {\bibinfo  {journal} {Prog. Part. Nucl. Phys.}\ }\textbf {\bibinfo {volume} {94}},\ \bibinfo {pages} {282} (\bibinfo {year} {2017})}\BibitemShut {NoStop}%
\bibitem [{\citenamefont {Esposito}\ \emph {et~al.}(2017)\citenamefont {Esposito}, \citenamefont {Pilloni},\ and\ \citenamefont {Polosa}}]{Esposito:2016noz}%
  \BibitemOpen
  \bibfield  {author} {\bibinfo {author} {\bibfnamefont {A.}~\bibnamefont {Esposito}}, \bibinfo {author} {\bibfnamefont {A.}~\bibnamefont {Pilloni}},\ and\ \bibinfo {author} {\bibfnamefont {A.~D.}\ \bibnamefont {Polosa}},\ }\bibfield  {title} {\bibinfo {title} {{Multiquark Resonances}},\ }\href {https://doi.org/10.1016/j.physrep.2016.11.002} {\bibfield  {journal} {\bibinfo  {journal} {Phys. Rept.}\ }\textbf {\bibinfo {volume} {668}},\ \bibinfo {pages} {1} (\bibinfo {year} {2017})},\ \Eprint {https://arxiv.org/abs/1611.07920} {arXiv:1611.07920 [hep-ph]} \BibitemShut {NoStop}%
\bibitem [{\citenamefont {Hosaka}\ \emph {et~al.}(2016)\citenamefont {Hosaka}, \citenamefont {Iijima}, \citenamefont {Miyabayashi}, \citenamefont {Sakai},\ and\ \citenamefont {Yasui}}]{Hosaka:2016pey}%
  \BibitemOpen
  \bibfield  {author} {\bibinfo {author} {\bibfnamefont {A.}~\bibnamefont {Hosaka}}, \bibinfo {author} {\bibfnamefont {T.}~\bibnamefont {Iijima}}, \bibinfo {author} {\bibfnamefont {K.}~\bibnamefont {Miyabayashi}}, \bibinfo {author} {\bibfnamefont {Y.}~\bibnamefont {Sakai}},\ and\ \bibinfo {author} {\bibfnamefont {S.}~\bibnamefont {Yasui}},\ }\bibfield  {title} {\bibinfo {title} {{Exotic hadrons with heavy flavors: X, Y, Z, and related states}},\ }\href {https://doi.org/10.1093/ptep/ptw045} {\bibfield  {journal} {\bibinfo  {journal} {PTEP}\ }\textbf {\bibinfo {volume} {2016}},\ \bibinfo {pages} {062C01} (\bibinfo {year} {2016})},\ \Eprint {https://arxiv.org/abs/1603.09229} {arXiv:1603.09229 [hep-ph]} \BibitemShut {NoStop}%
\bibitem [{\citenamefont {Lebed}\ \emph {et~al.}(2017)\citenamefont {Lebed}, \citenamefont {Mitchell},\ and\ \citenamefont {Swanson}}]{Lebed:2016hpi}%
  \BibitemOpen
  \bibfield  {author} {\bibinfo {author} {\bibfnamefont {R.~F.}\ \bibnamefont {Lebed}}, \bibinfo {author} {\bibfnamefont {R.~E.}\ \bibnamefont {Mitchell}},\ and\ \bibinfo {author} {\bibfnamefont {E.~S.}\ \bibnamefont {Swanson}},\ }\bibfield  {title} {\bibinfo {title} {{Heavy-Quark QCD Exotica}},\ }\href {https://doi.org/10.1016/j.ppnp.2016.11.003} {\bibfield  {journal} {\bibinfo  {journal} {Prog. Part. Nucl. Phys.}\ }\textbf {\bibinfo {volume} {93}},\ \bibinfo {pages} {143} (\bibinfo {year} {2017})},\ \Eprint {https://arxiv.org/abs/1610.04528} {arXiv:1610.04528 [hep-ph]} \BibitemShut {NoStop}%
\bibitem [{\citenamefont {Olsen}(2015)}]{Olsen:2014qna}%
  \BibitemOpen
  \bibfield  {author} {\bibinfo {author} {\bibfnamefont {S.~L.}\ \bibnamefont {Olsen}},\ }\bibfield  {title} {\bibinfo {title} {{A New Hadron Spectroscopy}},\ }\href {https://doi.org/10.1007/S11467-014-0449-6} {\bibfield  {journal} {\bibinfo  {journal} {Front. Phys. (Beijing)}\ }\textbf {\bibinfo {volume} {10}},\ \bibinfo {pages} {121} (\bibinfo {year} {2015})},\ \Eprint {https://arxiv.org/abs/1411.7738} {arXiv:1411.7738 [hep-ex]} \BibitemShut {NoStop}%
\bibitem [{\citenamefont {Olsen}\ \emph {et~al.}(2018)\citenamefont {Olsen}, \citenamefont {Skwarnicki},\ and\ \citenamefont {Zieminska}}]{Olsen:2017bmm}%
  \BibitemOpen
  \bibfield  {author} {\bibinfo {author} {\bibfnamefont {S.~L.}\ \bibnamefont {Olsen}}, \bibinfo {author} {\bibfnamefont {T.}~\bibnamefont {Skwarnicki}},\ and\ \bibinfo {author} {\bibfnamefont {D.}~\bibnamefont {Zieminska}},\ }\bibfield  {title} {\bibinfo {title} {{Nonstandard heavy mesons and baryons: Experimental evidence}},\ }\href {https://doi.org/10.1103/RevModPhys.90.015003} {\bibfield  {journal} {\bibinfo  {journal} {Rev. Mod. Phys.}\ }\textbf {\bibinfo {volume} {90}},\ \bibinfo {pages} {015003} (\bibinfo {year} {2018})},\ \Eprint {https://arxiv.org/abs/1708.04012} {arXiv:1708.04012 [hep-ph]} \BibitemShut {NoStop}%
\bibitem [{\citenamefont {Oset}\ \emph {et~al.}(2016)\citenamefont {Oset}, \citenamefont {Liang}, \citenamefont {Bayar} \emph {et~al.}}]{Oset:2016lyh}%
  \BibitemOpen
  \bibfield  {author} {\bibinfo {author} {\bibfnamefont {E.}~\bibnamefont {Oset}}, \bibinfo {author} {\bibfnamefont {W.-H.}\ \bibnamefont {Liang}}, \bibinfo {author} {\bibfnamefont {M.}~\bibnamefont {Bayar}}, \emph {et~al.},\ }\bibfield  {title} {\bibinfo {title} {{Weak decays of heavy hadrons into dynamically generated resonances}},\ }\href {https://doi.org/10.1142/S0218301316300010} {\bibfield  {journal} {\bibinfo  {journal} {Int. J. Mod. Phys. E}\ }\textbf {\bibinfo {volume} {25}},\ \bibinfo {pages} {1630001} (\bibinfo {year} {2016})},\ \Eprint {https://arxiv.org/abs/1601.03972} {arXiv:1601.03972 [hep-ph]} \BibitemShut {NoStop}%
\bibitem [{\citenamefont {Guo}\ \emph {et~al.}(2018)\citenamefont {Guo}, \citenamefont {Hanhart}, \citenamefont {Mei{\ss}ner}, \citenamefont {Wang}, \citenamefont {Zhao},\ and\ \citenamefont {Zou}}]{Guo:2017jvc}%
  \BibitemOpen
  \bibfield  {author} {\bibinfo {author} {\bibfnamefont {F.-K.}\ \bibnamefont {Guo}}, \bibinfo {author} {\bibfnamefont {C.}~\bibnamefont {Hanhart}}, \bibinfo {author} {\bibfnamefont {U.-G.}\ \bibnamefont {Mei{\ss}ner}}, \bibinfo {author} {\bibfnamefont {Q.}~\bibnamefont {Wang}}, \bibinfo {author} {\bibfnamefont {Q.}~\bibnamefont {Zhao}},\ and\ \bibinfo {author} {\bibfnamefont {B.-S.}\ \bibnamefont {Zou}},\ }\bibfield  {title} {\bibinfo {title} {{Hadronic molecules}},\ }\href {https://doi.org/10.1103/RevModPhys.90.015004} {\bibfield  {journal} {\bibinfo  {journal} {Rev. Mod. Phys.}\ }\textbf {\bibinfo {volume} {90}},\ \bibinfo {pages} {015004} (\bibinfo {year} {2018})},\ \bibinfo {note} {[Erratum: Rev.Mod.Phys. 94, 029901 (2022)]},\ \Eprint {https://arxiv.org/abs/1705.00141} {arXiv:1705.00141 [hep-ph]} \BibitemShut {NoStop}%
\bibitem [{\citenamefont {Lutz}\ and\ \citenamefont {Kolomeitsev}(2004)}]{Lutz:2003fm}%
  \BibitemOpen
  \bibfield  {author} {\bibinfo {author} {\bibfnamefont {M.~F.~M.}\ \bibnamefont {Lutz}}\ and\ \bibinfo {author} {\bibfnamefont {E.~E.}\ \bibnamefont {Kolomeitsev}},\ }\bibfield  {title} {\bibinfo {title} {{On meson resonances and chiral symmetry}},\ }\href {https://doi.org/10.1016/j.nuclphysa.2003.11.009} {\bibfield  {journal} {\bibinfo  {journal} {Nucl. Phys. A}\ }\textbf {\bibinfo {volume} {730}},\ \bibinfo {pages} {392} (\bibinfo {year} {2004})},\ \Eprint {https://arxiv.org/abs/nucl-th/0307039} {arXiv:nucl-th/0307039} \BibitemShut {NoStop}%
\bibitem [{\citenamefont {Roca}\ \emph {et~al.}(2005)\citenamefont {Roca}, \citenamefont {Oset},\ and\ \citenamefont {Singh}}]{Roca:2005nm}%
  \BibitemOpen
  \bibfield  {author} {\bibinfo {author} {\bibfnamefont {L.}~\bibnamefont {Roca}}, \bibinfo {author} {\bibfnamefont {E.}~\bibnamefont {Oset}},\ and\ \bibinfo {author} {\bibfnamefont {J.}~\bibnamefont {Singh}},\ }\bibfield  {title} {\bibinfo {title} {{Low lying axial-vector mesons as dynamically generated resonances}},\ }\href {https://doi.org/10.1103/PhysRevD.72.014002} {\bibfield  {journal} {\bibinfo  {journal} {Phys. Rev. D}\ }\textbf {\bibinfo {volume} {72}},\ \bibinfo {pages} {014002} (\bibinfo {year} {2005})},\ \Eprint {https://arxiv.org/abs/hep-ph/0503273} {arXiv:hep-ph/0503273} \BibitemShut {NoStop}%
\bibitem [{\citenamefont {Garcia-Recio}\ \emph {et~al.}(2011)\citenamefont {Garcia-Recio}, \citenamefont {Geng}, \citenamefont {Nieves},\ and\ \citenamefont {Salcedo}}]{Garcia-Recio:2010enl}%
  \BibitemOpen
  \bibfield  {author} {\bibinfo {author} {\bibfnamefont {C.}~\bibnamefont {Garcia-Recio}}, \bibinfo {author} {\bibfnamefont {L.~S.}\ \bibnamefont {Geng}}, \bibinfo {author} {\bibfnamefont {J.}~\bibnamefont {Nieves}},\ and\ \bibinfo {author} {\bibfnamefont {L.~L.}\ \bibnamefont {Salcedo}},\ }\bibfield  {title} {\bibinfo {title} {{Low-lying even parity meson resonances and spin-flavor symmetry}},\ }\href {https://doi.org/10.1103/PhysRevD.83.016007} {\bibfield  {journal} {\bibinfo  {journal} {Phys. Rev. D}\ }\textbf {\bibinfo {volume} {83}},\ \bibinfo {pages} {016007} (\bibinfo {year} {2011})},\ \Eprint {https://arxiv.org/abs/1005.0956} {arXiv:1005.0956 [hep-ph]} \BibitemShut {NoStop}%
\bibitem [{\citenamefont {Zhou}\ \emph {et~al.}(2014)\citenamefont {Zhou}, \citenamefont {Ren}, \citenamefont {Chen},\ and\ \citenamefont {Geng}}]{Zhou:2014ila}%
  \BibitemOpen
  \bibfield  {author} {\bibinfo {author} {\bibfnamefont {Y.}~\bibnamefont {Zhou}}, \bibinfo {author} {\bibfnamefont {X.-L.}\ \bibnamefont {Ren}}, \bibinfo {author} {\bibfnamefont {H.-X.}\ \bibnamefont {Chen}},\ and\ \bibinfo {author} {\bibfnamefont {L.-S.}\ \bibnamefont {Geng}},\ }\bibfield  {title} {\bibinfo {title} {{Pseudoscalar meson and vector meson interactions and dynamically generated axial-vector mesons}},\ }\href {https://doi.org/10.1103/PhysRevD.90.014020} {\bibfield  {journal} {\bibinfo  {journal} {Phys. Rev. D}\ }\textbf {\bibinfo {volume} {90}},\ \bibinfo {pages} {014020} (\bibinfo {year} {2014})},\ \Eprint {https://arxiv.org/abs/1404.6847} {arXiv:1404.6847 [nucl-th]} \BibitemShut {NoStop}%
\bibitem [{\citenamefont {Geng}\ \emph {et~al.}(2015)\citenamefont {Geng}, \citenamefont {Ren}, \citenamefont {Zhou}, \citenamefont {Chen},\ and\ \citenamefont {Oset}}]{Geng:2015yta}%
  \BibitemOpen
  \bibfield  {author} {\bibinfo {author} {\bibfnamefont {L.-S.}\ \bibnamefont {Geng}}, \bibinfo {author} {\bibfnamefont {X.-L.}\ \bibnamefont {Ren}}, \bibinfo {author} {\bibfnamefont {Y.}~\bibnamefont {Zhou}}, \bibinfo {author} {\bibfnamefont {H.-X.}\ \bibnamefont {Chen}},\ and\ \bibinfo {author} {\bibfnamefont {E.}~\bibnamefont {Oset}},\ }\bibfield  {title} {\bibinfo {title} {{$S$-wave $KK^*$ interactions in a finite volume and the $f_1(1285)$}},\ }\href {https://doi.org/10.1103/PhysRevD.92.014029} {\bibfield  {journal} {\bibinfo  {journal} {Phys. Rev. D}\ }\textbf {\bibinfo {volume} {92}},\ \bibinfo {pages} {014029} (\bibinfo {year} {2015})},\ \Eprint {https://arxiv.org/abs/1503.06633} {arXiv:1503.06633 [hep-ph]} \BibitemShut {NoStop}%
\bibitem [{\citenamefont {L{\"u}}\ and\ \citenamefont {He}(2016)}]{Lu:2016nlp}%
  \BibitemOpen
  \bibfield  {author} {\bibinfo {author} {\bibfnamefont {P.-L.}\ \bibnamefont {L{\"u}}}\ and\ \bibinfo {author} {\bibfnamefont {J.}~\bibnamefont {He}},\ }\bibfield  {title} {\bibinfo {title} {{Hadronic molecular states from the $K\bar{K}^*$ interaction}},\ }\href {https://doi.org/10.1140/epja/i2016-16359-7} {\bibfield  {journal} {\bibinfo  {journal} {Eur. Phys. J. A}\ }\textbf {\bibinfo {volume} {52}},\ \bibinfo {pages} {359} (\bibinfo {year} {2016})},\ \Eprint {https://arxiv.org/abs/1603.04168} {arXiv:1603.04168 [hep-ph]} \BibitemShut {NoStop}%
\bibitem [{\citenamefont {Jia}\ \emph {et~al.}(2026{\natexlab{a}})\citenamefont {Jia}, \citenamefont {Song}, \citenamefont {Liang},\ and\ \citenamefont {Oset}}]{Jia:2026dpl}%
  \BibitemOpen
  \bibfield  {author} {\bibinfo {author} {\bibfnamefont {W.-H.}\ \bibnamefont {Jia}}, \bibinfo {author} {\bibfnamefont {J.}~\bibnamefont {Song}}, \bibinfo {author} {\bibfnamefont {W.-H.}\ \bibnamefont {Liang}},\ and\ \bibinfo {author} {\bibfnamefont {E.}~\bibnamefont {Oset}},\ }\href@noop {} {\bibinfo {title} {{Scattering data and correlation function for the $K f_1(1285)$ interaction}}} (\bibinfo {year} {2026}{\natexlab{a}}),\ \Eprint {https://arxiv.org/abs/2602.16683} {arXiv:2602.16683 [hep-ph]} \BibitemShut {NoStop}%
\bibitem [{\citenamefont {Encarnaci{\'o}n}\ \emph {et~al.}(2025)\citenamefont {Encarnaci{\'o}n}, \citenamefont {Feijoo},\ and\ \citenamefont {Oset}}]{Encarnacion:2025lyf}%
  \BibitemOpen
  \bibfield  {author} {\bibinfo {author} {\bibfnamefont {P.}~\bibnamefont {Encarnaci{\'o}n}}, \bibinfo {author} {\bibfnamefont {A.}~\bibnamefont {Feijoo}},\ and\ \bibinfo {author} {\bibfnamefont {E.}~\bibnamefont {Oset}},\ }\bibfield  {title} {\bibinfo {title} {{Correlation function for the $p f_1(1285)$ interaction}},\ }\href {https://doi.org/10.1103/s7g4-6lmv} {\bibfield  {journal} {\bibinfo  {journal} {Phys. Rev. D}\ }\textbf {\bibinfo {volume} {111}},\ \bibinfo {pages} {114023} (\bibinfo {year} {2025})},\ \Eprint {https://arxiv.org/abs/2502.19329} {arXiv:2502.19329 [hep-ph]} \BibitemShut {NoStop}%
\bibitem [{\citenamefont {Kamalov}\ \emph {et~al.}(2001)\citenamefont {Kamalov}, \citenamefont {Oset},\ and\ \citenamefont {Ramos}}]{Kamalov:2000iy}%
  \BibitemOpen
  \bibfield  {author} {\bibinfo {author} {\bibfnamefont {S.~S.}\ \bibnamefont {Kamalov}}, \bibinfo {author} {\bibfnamefont {E.}~\bibnamefont {Oset}},\ and\ \bibinfo {author} {\bibfnamefont {A.}~\bibnamefont {Ramos}},\ }\bibfield  {title} {\bibinfo {title} {{Chiral unitary approach to the $K^-$ deuteron scattering length}},\ }\href {https://doi.org/10.1016/S0375-9474(00)00709-0} {\bibfield  {journal} {\bibinfo  {journal} {Nucl. Phys. A}\ }\textbf {\bibinfo {volume} {690}},\ \bibinfo {pages} {494} (\bibinfo {year} {2001})},\ \Eprint {https://arxiv.org/abs/nucl-th/0010054} {arXiv:nucl-th/0010054} \BibitemShut {NoStop}%
\bibitem [{\citenamefont {Foldy}(1945)}]{Foldy:1945zz}%
  \BibitemOpen
  \bibfield  {author} {\bibinfo {author} {\bibfnamefont {L.~L.}\ \bibnamefont {Foldy}},\ }\bibfield  {title} {\bibinfo {title} {{The Multiple Scattering of Waves. 1. General Theory of Isotropic Scattering by Randomly Distributed Scatterers}},\ }\href {https://doi.org/10.1103/PhysRev.67.107} {\bibfield  {journal} {\bibinfo  {journal} {Phys. Rev.}\ }\textbf {\bibinfo {volume} {67}},\ \bibinfo {pages} {107} (\bibinfo {year} {1945})}\BibitemShut {NoStop}%
\bibitem [{\citenamefont {Brueckner}(1953{\natexlab{a}})}]{Brueckner:1953zz}%
  \BibitemOpen
  \bibfield  {author} {\bibinfo {author} {\bibfnamefont {K.~A.}\ \bibnamefont {Brueckner}},\ }\bibfield  {title} {\bibinfo {title} {{Multiple Scattering Corrections to the Impulse Approximation in the Two-Body System}},\ }\href {https://doi.org/10.1103/PhysRev.89.834} {\bibfield  {journal} {\bibinfo  {journal} {Phys. Rev.}\ }\textbf {\bibinfo {volume} {89}},\ \bibinfo {pages} {834} (\bibinfo {year} {1953}{\natexlab{a}})}\BibitemShut {NoStop}%
\bibitem [{\citenamefont {Brueckner}(1953{\natexlab{b}})}]{Brueckner:1953zza}%
  \BibitemOpen
  \bibfield  {author} {\bibinfo {author} {\bibfnamefont {K.~A.}\ \bibnamefont {Brueckner}},\ }\bibfield  {title} {\bibinfo {title} {{The Elastic Scattering of Pions in Deuterium}},\ }\href {https://doi.org/10.1103/PhysRev.90.715} {\bibfield  {journal} {\bibinfo  {journal} {Phys. Rev.}\ }\textbf {\bibinfo {volume} {90}},\ \bibinfo {pages} {715} (\bibinfo {year} {1953}{\natexlab{b}})}\BibitemShut {NoStop}%
\bibitem [{\citenamefont {Chand}\ and\ \citenamefont {Dalitz}(1962)}]{Chand:1962ec}%
  \BibitemOpen
  \bibfield  {author} {\bibinfo {author} {\bibfnamefont {R.}~\bibnamefont {Chand}}\ and\ \bibinfo {author} {\bibfnamefont {R.~H.}\ \bibnamefont {Dalitz}},\ }\bibfield  {title} {\bibinfo {title} {{Charge-independence in $K^-$-deuterium capture reactions}},\ }\href {https://doi.org/10.1016/0003-4916(62)90113-6} {\bibfield  {journal} {\bibinfo  {journal} {Annals Phys.}\ }\textbf {\bibinfo {volume} {20}},\ \bibinfo {pages} {1} (\bibinfo {year} {1962})}\BibitemShut {NoStop}%
\bibitem [{\citenamefont {Barrett}\ and\ \citenamefont {Deloff}(1999)}]{Barrett:1999cw}%
  \BibitemOpen
  \bibfield  {author} {\bibinfo {author} {\bibfnamefont {R.~C.}\ \bibnamefont {Barrett}}\ and\ \bibinfo {author} {\bibfnamefont {A.}~\bibnamefont {Deloff}},\ }\bibfield  {title} {\bibinfo {title} {{Strong interaction effects in kaonic deuterium}},\ }\href {https://doi.org/10.1103/PhysRevC.60.025201} {\bibfield  {journal} {\bibinfo  {journal} {Phys. Rev. C}\ }\textbf {\bibinfo {volume} {60}},\ \bibinfo {pages} {025201} (\bibinfo {year} {1999})}\BibitemShut {NoStop}%
\bibitem [{\citenamefont {Deloff}(2000)}]{Deloff:1999gc}%
  \BibitemOpen
  \bibfield  {author} {\bibinfo {author} {\bibfnamefont {A.}~\bibnamefont {Deloff}},\ }\bibfield  {title} {\bibinfo {title} {{$\eta d$ and $K^- d$ zero energy scattering: A Faddeev approach}},\ }\href {https://doi.org/10.1103/PhysRevC.61.024004} {\bibfield  {journal} {\bibinfo  {journal} {Phys. Rev. C}\ }\textbf {\bibinfo {volume} {61}},\ \bibinfo {pages} {024004} (\bibinfo {year} {2000})}\BibitemShut {NoStop}%
\bibitem [{\citenamefont {Martinez~Torres}\ \emph {et~al.}(2020)\citenamefont {Martinez~Torres}, \citenamefont {Khemchandani}, \citenamefont {Roca},\ and\ \citenamefont {Oset}}]{MartinezTorres:2020hus}%
  \BibitemOpen
  \bibfield  {author} {\bibinfo {author} {\bibfnamefont {A.}~\bibnamefont {Martinez~Torres}}, \bibinfo {author} {\bibfnamefont {K.~P.}\ \bibnamefont {Khemchandani}}, \bibinfo {author} {\bibfnamefont {L.}~\bibnamefont {Roca}},\ and\ \bibinfo {author} {\bibfnamefont {E.}~\bibnamefont {Oset}},\ }\bibfield  {title} {\bibinfo {title} {{Few-body systems consisting of mesons}},\ }\href {https://doi.org/10.1007/s00601-020-01568-y} {\bibfield  {journal} {\bibinfo  {journal} {Few Body Syst.}\ }\textbf {\bibinfo {volume} {61}},\ \bibinfo {pages} {35} (\bibinfo {year} {2020})},\ \Eprint {https://arxiv.org/abs/2005.14357} {arXiv:2005.14357 [nucl-th]} \BibitemShut {NoStop}%
\bibitem [{\citenamefont {Roca}\ and\ \citenamefont {Oset}(2010)}]{Roca:2010tf}%
  \BibitemOpen
  \bibfield  {author} {\bibinfo {author} {\bibfnamefont {L.}~\bibnamefont {Roca}}\ and\ \bibinfo {author} {\bibfnamefont {E.}~\bibnamefont {Oset}},\ }\bibfield  {title} {\bibinfo {title} {{A description of the $f_2(1270)$, $\rho_3(1690)$, $f_4(2050)$, $\rho_5(2350)$ and $f_6(2510)$ resonances as multi-$\rho(770)$ states}},\ }\href {https://doi.org/10.1103/PhysRevD.82.054013} {\bibfield  {journal} {\bibinfo  {journal} {Phys. Rev. D}\ }\textbf {\bibinfo {volume} {82}},\ \bibinfo {pages} {054013} (\bibinfo {year} {2010})},\ \Eprint {https://arxiv.org/abs/1005.0283} {arXiv:1005.0283 [hep-ph]} \BibitemShut {NoStop}%
\bibitem [{\citenamefont {Malabarba}\ \emph {et~al.}(2025)\citenamefont {Malabarba}, \citenamefont {Khemchandani}, \citenamefont {Martinez~Torres},\ and\ \citenamefont {Nam}}]{Malabarba:2024hlv}%
  \BibitemOpen
  \bibfield  {author} {\bibinfo {author} {\bibfnamefont {B.~B.}\ \bibnamefont {Malabarba}}, \bibinfo {author} {\bibfnamefont {K.~P.}\ \bibnamefont {Khemchandani}}, \bibinfo {author} {\bibfnamefont {A.}~\bibnamefont {Martinez~Torres}},\ and\ \bibinfo {author} {\bibfnamefont {S.-i.}\ \bibnamefont {Nam}},\ }\bibfield  {title} {\bibinfo {title} {{Strangeness +1 light multiquark baryons}},\ }\href {https://doi.org/10.1103/PhysRevD.111.016021} {\bibfield  {journal} {\bibinfo  {journal} {Phys. Rev. D}\ }\textbf {\bibinfo {volume} {111}},\ \bibinfo {pages} {016021} (\bibinfo {year} {2025})},\ \Eprint {https://arxiv.org/abs/2404.04078} {arXiv:2404.04078 [hep-ph]} \BibitemShut {NoStop}%
\bibitem [{\citenamefont {Ericson}\ and\ \citenamefont {Weise}(1988)}]{Ericson:1988gk}%
  \BibitemOpen
  \bibfield  {author} {\bibinfo {author} {\bibfnamefont {T.~E.~O.}\ \bibnamefont {Ericson}}\ and\ \bibinfo {author} {\bibfnamefont {W.}~\bibnamefont {Weise}},\ }\href@noop {} {\emph {\bibinfo {title} {{Pions and Nuclei}}}}\ (\bibinfo  {publisher} {Clarendon Press},\ \bibinfo {address} {Oxford, UK},\ \bibinfo {year} {1988})\BibitemShut {NoStop}%
\bibitem [{\citenamefont {Seki}\ and\ \citenamefont {Masutani}(1983)}]{Seki:1983sh}%
  \BibitemOpen
  \bibfield  {author} {\bibinfo {author} {\bibfnamefont {R.}~\bibnamefont {Seki}}\ and\ \bibinfo {author} {\bibfnamefont {K.}~\bibnamefont {Masutani}},\ }\bibfield  {title} {\bibinfo {title} {{Unified analysis of pionic atoms and low-energy pion-nucleus scattering: Phenomenological analysis}},\ }\href {https://doi.org/10.1103/PhysRevC.27.2799} {\bibfield  {journal} {\bibinfo  {journal} {Phys. Rev. C}\ }\textbf {\bibinfo {volume} {27}},\ \bibinfo {pages} {2799} (\bibinfo {year} {1983})}\BibitemShut {NoStop}%
\bibitem [{\citenamefont {Nieves}\ \emph {et~al.}(1993)\citenamefont {Nieves}, \citenamefont {Oset},\ and\ \citenamefont {Garcia-Recio}}]{Nieves:1993ev}%
  \BibitemOpen
  \bibfield  {author} {\bibinfo {author} {\bibfnamefont {J.}~\bibnamefont {Nieves}}, \bibinfo {author} {\bibfnamefont {E.}~\bibnamefont {Oset}},\ and\ \bibinfo {author} {\bibfnamefont {C.}~\bibnamefont {Garcia-Recio}},\ }\bibfield  {title} {\bibinfo {title} {{A Theoretical approach to pionic atoms and the problem of anomalies}},\ }\href {https://doi.org/10.1016/0375-9474(93)90245-S} {\bibfield  {journal} {\bibinfo  {journal} {Nucl. Phys. A}\ }\textbf {\bibinfo {volume} {554}},\ \bibinfo {pages} {509} (\bibinfo {year} {1993})}\BibitemShut {NoStop}%
\bibitem [{\citenamefont {Brown}\ and\ \citenamefont {Weise}(1975)}]{Brown:1975di}%
  \BibitemOpen
  \bibfield  {author} {\bibinfo {author} {\bibfnamefont {G.~E.}\ \bibnamefont {Brown}}\ and\ \bibinfo {author} {\bibfnamefont {W.}~\bibnamefont {Weise}},\ }\bibfield  {title} {\bibinfo {title} {{Pion Scattering and Isobars in Nuclei}},\ }\href {https://doi.org/10.1016/0370-1573(75)90026-5} {\bibfield  {journal} {\bibinfo  {journal} {Phys. Rept.}\ }\textbf {\bibinfo {volume} {22}},\ \bibinfo {pages} {279} (\bibinfo {year} {1975})}\BibitemShut {NoStop}%
\bibitem [{\citenamefont {Ikeno}\ and\ \citenamefont {Oset}(2025)}]{Ikeno:2025bsx}%
  \BibitemOpen
  \bibfield  {author} {\bibinfo {author} {\bibfnamefont {N.}~\bibnamefont {Ikeno}}\ and\ \bibinfo {author} {\bibfnamefont {E.}~\bibnamefont {Oset}},\ }\bibfield  {title} {\bibinfo {title} {{Correlation function for the $n \bar D_{s0}^*(2317)$ interaction and the issue of elastic unitarity}},\ }\href {https://doi.org/10.1103/bb31-rdjb} {\bibfield  {journal} {\bibinfo  {journal} {Phys. Rev. D}\ }\textbf {\bibinfo {volume} {112}},\ \bibinfo {pages} {094019} (\bibinfo {year} {2025})},\ \Eprint {https://arxiv.org/abs/2507.16367} {arXiv:2507.16367 [hep-ph]} \BibitemShut {NoStop}%
\bibitem [{\citenamefont {Yan}\ \emph {et~al.}(2023)\citenamefont {Yan}, \citenamefont {Dias}, \citenamefont {Guevara}, \citenamefont {Guo},\ and\ \citenamefont {Zou}}]{Yan:2023vbh}%
  \BibitemOpen
  \bibfield  {author} {\bibinfo {author} {\bibfnamefont {M.-J.}\ \bibnamefont {Yan}}, \bibinfo {author} {\bibfnamefont {J.~M.}\ \bibnamefont {Dias}}, \bibinfo {author} {\bibfnamefont {A.}~\bibnamefont {Guevara}}, \bibinfo {author} {\bibfnamefont {F.-K.}\ \bibnamefont {Guo}},\ and\ \bibinfo {author} {\bibfnamefont {B.-S.}\ \bibnamefont {Zou}},\ }\bibfield  {title} {\bibinfo {title} {{On the $\eta_1(1855)$, $\pi_1(1400)$ and $\pi_1(1600)$ as dynamically generated states and their SU(3) partners}},\ }\href {https://doi.org/10.3390/universe9020109} {\bibfield  {journal} {\bibinfo  {journal} {Universe}\ }\textbf {\bibinfo {volume} {9}},\ \bibinfo {pages} {109} (\bibinfo {year} {2023})},\ \Eprint {https://arxiv.org/abs/2301.04432} {arXiv:2301.04432 [hep-ph]} \BibitemShut {NoStop}%
\bibitem [{\citenamefont {Agat{\~a}o}\ \emph {et~al.}(2025)\citenamefont {Agat{\~a}o}, \citenamefont {Brand{\~a}o}, \citenamefont {Mart{\'\i}nez~Torres}, \citenamefont {Khemchandani}, \citenamefont {Abreu},\ and\ \citenamefont {Oset}}]{Agatao:2025ckp}%
  \BibitemOpen
  \bibfield  {author} {\bibinfo {author} {\bibfnamefont {B.}~\bibnamefont {Agat{\~a}o}}, \bibinfo {author} {\bibfnamefont {P.}~\bibnamefont {Brand{\~a}o}}, \bibinfo {author} {\bibfnamefont {A.}~\bibnamefont {Mart{\'\i}nez~Torres}}, \bibinfo {author} {\bibfnamefont {K.~P.}\ \bibnamefont {Khemchandani}}, \bibinfo {author} {\bibfnamefont {L.~M.}\ \bibnamefont {Abreu}},\ and\ \bibinfo {author} {\bibfnamefont {E.}~\bibnamefont {Oset}},\ }\bibfield  {title} {\bibinfo {title} {{Correlation functions for $n\,\bar{D}_{s1}(2460)$ and $n\,\bar{D}_{s1}(2536)$}},\ }\href {https://doi.org/10.1140/epjc/s10052-025-14838-y} {\bibfield  {journal} {\bibinfo  {journal} {Eur. Phys. J. C}\ }\textbf {\bibinfo {volume} {85}},\ \bibinfo {pages} {1136} (\bibinfo {year} {2025})},\ \Eprint {https://arxiv.org/abs/2508.05825} {arXiv:2508.05825 [hep-ph]} \BibitemShut {NoStop}%
\bibitem [{\citenamefont {Jia}\ \emph {et~al.}(2026{\natexlab{b}})\citenamefont {Jia}, \citenamefont {Su}, \citenamefont {Liang}, \citenamefont {Molina},\ and\ \citenamefont {Oset}}]{Jia:2025obs}%
  \BibitemOpen
  \bibfield  {author} {\bibinfo {author} {\bibfnamefont {W.-H.}\ \bibnamefont {Jia}}, \bibinfo {author} {\bibfnamefont {P.-S.}\ \bibnamefont {Su}}, \bibinfo {author} {\bibfnamefont {W.-H.}\ \bibnamefont {Liang}}, \bibinfo {author} {\bibfnamefont {R.}~\bibnamefont {Molina}},\ and\ \bibinfo {author} {\bibfnamefont {E.}~\bibnamefont {Oset}},\ }\bibfield  {title} {\bibinfo {title} {{Superexotic $K^{*+}D^{*+}K^{*+}$ bound state}},\ }\href {https://doi.org/10.1016/j.physletb.2026.140320} {\bibfield  {journal} {\bibinfo  {journal} {Phys. Lett. B}\ }\textbf {\bibinfo {volume} {875}},\ \bibinfo {pages} {140320} (\bibinfo {year} {2026}{\natexlab{b}})},\ \Eprint {https://arxiv.org/abs/2512.04001} {arXiv:2512.04001 [hep-ph]} \BibitemShut {NoStop}%
\bibitem [{\citenamefont {van Beveren}\ and\ \citenamefont {Rupp}(2003)}]{vanBeveren:2003kd}%
  \BibitemOpen
  \bibfield  {author} {\bibinfo {author} {\bibfnamefont {E.}~\bibnamefont {van Beveren}}\ and\ \bibinfo {author} {\bibfnamefont {G.}~\bibnamefont {Rupp}},\ }\bibfield  {title} {\bibinfo {title} {{Observed $D_s(2317)$ and tentative $D(2100-2300)$ as the charmed cousins of the light scalar nonet}},\ }\href {https://doi.org/10.1103/PhysRevLett.91.012003} {\bibfield  {journal} {\bibinfo  {journal} {Phys. Rev. Lett.}\ }\textbf {\bibinfo {volume} {91}},\ \bibinfo {pages} {012003} (\bibinfo {year} {2003})},\ \Eprint {https://arxiv.org/abs/hep-ph/0305035} {arXiv:hep-ph/0305035} \BibitemShut {NoStop}%
\bibitem [{\citenamefont {Barnes}\ \emph {et~al.}(2003)\citenamefont {Barnes}, \citenamefont {Close},\ and\ \citenamefont {Lipkin}}]{Barnes:2003dj}%
  \BibitemOpen
  \bibfield  {author} {\bibinfo {author} {\bibfnamefont {T.}~\bibnamefont {Barnes}}, \bibinfo {author} {\bibfnamefont {F.~E.}\ \bibnamefont {Close}},\ and\ \bibinfo {author} {\bibfnamefont {H.~J.}\ \bibnamefont {Lipkin}},\ }\bibfield  {title} {\bibinfo {title} {{Implications of a $DK$ molecule at 2.32 GeV}},\ }\href {https://doi.org/10.1103/PhysRevD.68.054006} {\bibfield  {journal} {\bibinfo  {journal} {Phys. Rev. D}\ }\textbf {\bibinfo {volume} {68}},\ \bibinfo {pages} {054006} (\bibinfo {year} {2003})},\ \Eprint {https://arxiv.org/abs/hep-ph/0305025} {arXiv:hep-ph/0305025} \BibitemShut {NoStop}%
\bibitem [{\citenamefont {Chen}\ and\ \citenamefont {Li}(2004)}]{Chen:2004dy}%
  \BibitemOpen
  \bibfield  {author} {\bibinfo {author} {\bibfnamefont {Y.-Q.}\ \bibnamefont {Chen}}\ and\ \bibinfo {author} {\bibfnamefont {X.-Q.}\ \bibnamefont {Li}},\ }\bibfield  {title} {\bibinfo {title} {{A Comprehensive four-quark interpretation of $D_s(2317)$, $D_s(2457)$ and $D_s(2632)$}},\ }\href {https://doi.org/10.1103/PhysRevLett.93.232001} {\bibfield  {journal} {\bibinfo  {journal} {Phys. Rev. Lett.}\ }\textbf {\bibinfo {volume} {93}},\ \bibinfo {pages} {232001} (\bibinfo {year} {2004})},\ \Eprint {https://arxiv.org/abs/hep-ph/0407062} {arXiv:hep-ph/0407062} \BibitemShut {NoStop}%
\bibitem [{\citenamefont {Kolomeitsev}\ and\ \citenamefont {Lutz}(2004)}]{Kolomeitsev:2003ac}%
  \BibitemOpen
  \bibfield  {author} {\bibinfo {author} {\bibfnamefont {E.~E.}\ \bibnamefont {Kolomeitsev}}\ and\ \bibinfo {author} {\bibfnamefont {M.~F.~M.}\ \bibnamefont {Lutz}},\ }\bibfield  {title} {\bibinfo {title} {{On Heavy light meson resonances and chiral symmetry}},\ }\href {https://doi.org/10.1016/j.physletb.2003.10.118} {\bibfield  {journal} {\bibinfo  {journal} {Phys. Lett. B}\ }\textbf {\bibinfo {volume} {582}},\ \bibinfo {pages} {39} (\bibinfo {year} {2004})},\ \Eprint {https://arxiv.org/abs/hep-ph/0307133} {arXiv:hep-ph/0307133} \BibitemShut {NoStop}%
\bibitem [{\citenamefont {Gamermann}\ \emph {et~al.}(2007)\citenamefont {Gamermann}, \citenamefont {Oset}, \citenamefont {Strottman},\ and\ \citenamefont {Vicente~Vacas}}]{Gamermann:2006nm}%
  \BibitemOpen
  \bibfield  {author} {\bibinfo {author} {\bibfnamefont {D.}~\bibnamefont {Gamermann}}, \bibinfo {author} {\bibfnamefont {E.}~\bibnamefont {Oset}}, \bibinfo {author} {\bibfnamefont {D.}~\bibnamefont {Strottman}},\ and\ \bibinfo {author} {\bibfnamefont {M.~J.}\ \bibnamefont {Vicente~Vacas}},\ }\bibfield  {title} {\bibinfo {title} {{Dynamically generated open and hidden charm meson systems}},\ }\href {https://doi.org/10.1103/PhysRevD.76.074016} {\bibfield  {journal} {\bibinfo  {journal} {Phys. Rev. D}\ }\textbf {\bibinfo {volume} {76}},\ \bibinfo {pages} {074016} (\bibinfo {year} {2007})},\ \Eprint {https://arxiv.org/abs/hep-ph/0612179} {arXiv:hep-ph/0612179} \BibitemShut {NoStop}%
\bibitem [{\citenamefont {Guo}\ \emph {et~al.}(2007)\citenamefont {Guo}, \citenamefont {Shen},\ and\ \citenamefont {Chiang}}]{Guo:2006rp}%
  \BibitemOpen
  \bibfield  {author} {\bibinfo {author} {\bibfnamefont {F.-K.}\ \bibnamefont {Guo}}, \bibinfo {author} {\bibfnamefont {P.-N.}\ \bibnamefont {Shen}},\ and\ \bibinfo {author} {\bibfnamefont {H.-C.}\ \bibnamefont {Chiang}},\ }\bibfield  {title} {\bibinfo {title} {{Dynamically generated $1^+$ heavy mesons}},\ }\href {https://doi.org/10.1016/j.physletb.2007.01.050} {\bibfield  {journal} {\bibinfo  {journal} {Phys. Lett. B}\ }\textbf {\bibinfo {volume} {647}},\ \bibinfo {pages} {133} (\bibinfo {year} {2007})},\ \Eprint {https://arxiv.org/abs/hep-ph/0610008} {arXiv:hep-ph/0610008} \BibitemShut {NoStop}%
\bibitem [{\citenamefont {Yang}\ \emph {et~al.}(2022)\citenamefont {Yang}, \citenamefont {Wang}, \citenamefont {Wu}, \citenamefont {Oka},\ and\ \citenamefont {Zhu}}]{Yang:2021tvc}%
  \BibitemOpen
  \bibfield  {author} {\bibinfo {author} {\bibfnamefont {Z.}~\bibnamefont {Yang}}, \bibinfo {author} {\bibfnamefont {G.-J.}\ \bibnamefont {Wang}}, \bibinfo {author} {\bibfnamefont {J.-J.}\ \bibnamefont {Wu}}, \bibinfo {author} {\bibfnamefont {M.}~\bibnamefont {Oka}},\ and\ \bibinfo {author} {\bibfnamefont {S.-L.}\ \bibnamefont {Zhu}},\ }\bibfield  {title} {\bibinfo {title} {{Novel Coupled Channel Framework Connecting the Quark Model and Lattice QCD for the Near-threshold Ds States}},\ }\href {https://doi.org/10.1103/PhysRevLett.128.112001} {\bibfield  {journal} {\bibinfo  {journal} {Phys. Rev. Lett.}\ }\textbf {\bibinfo {volume} {128}},\ \bibinfo {pages} {112001} (\bibinfo {year} {2022})},\ \Eprint {https://arxiv.org/abs/2107.04860} {arXiv:2107.04860 [hep-ph]} \BibitemShut {NoStop}%
\bibitem [{\citenamefont {Liu}\ \emph {et~al.}(2022)\citenamefont {Liu}, \citenamefont {Ling}, \citenamefont {Geng}, \citenamefont {En-Wang},\ and\ \citenamefont {Xie}}]{Liu:2022dmm}%
  \BibitemOpen
  \bibfield  {author} {\bibinfo {author} {\bibfnamefont {M.-Z.}\ \bibnamefont {Liu}}, \bibinfo {author} {\bibfnamefont {X.-Z.}\ \bibnamefont {Ling}}, \bibinfo {author} {\bibfnamefont {L.-S.}\ \bibnamefont {Geng}}, \bibinfo {author} {\bibnamefont {En-Wang}},\ and\ \bibinfo {author} {\bibfnamefont {J.-J.}\ \bibnamefont {Xie}},\ }\bibfield  {title} {\bibinfo {title} {{Production of $D_{s0}^*(2317)$ and $D_{s1}(2460)$ in $B$ decays as $D^{(*)}K$ and $D_s^{(*)}\eta$ molecules}},\ }\href {https://doi.org/10.1103/PhysRevD.106.114011} {\bibfield  {journal} {\bibinfo  {journal} {Phys. Rev. D}\ }\textbf {\bibinfo {volume} {106}},\ \bibinfo {pages} {114011} (\bibinfo {year} {2022})},\ \Eprint {https://arxiv.org/abs/2209.01103} {arXiv:2209.01103 [hep-ph]} \BibitemShut {NoStop}%
\bibitem [{\citenamefont {Liang}\ \emph {et~al.}(2025)\citenamefont {Liang}, \citenamefont {Hu},\ and\ \citenamefont {Oset}}]{Liang:2025jkj}%
  \BibitemOpen
  \bibfield  {author} {\bibinfo {author} {\bibfnamefont {W.-H.}\ \bibnamefont {Liang}}, \bibinfo {author} {\bibfnamefont {Z.-R.}\ \bibnamefont {Hu}},\ and\ \bibinfo {author} {\bibfnamefont {E.}~\bibnamefont {Oset}},\ }\href@noop {} {\bibinfo {title} {{The $B^{+(0)} \to \bar D^{0(-)} D^{*}_{s0}(2317)^+$ decays and the molecular structure of $D^*_{s0}(2317)$}}} (\bibinfo {year} {2025}),\ \Eprint {https://arxiv.org/abs/2512.22604} {arXiv:2512.22604 [hep-ph]} \BibitemShut {NoStop}%
\bibitem [{\citenamefont {Mohler}\ \emph {et~al.}(2013)\citenamefont {Mohler}, \citenamefont {Lang}, \citenamefont {Leskovec}, \citenamefont {Prelovsek},\ and\ \citenamefont {Woloshyn}}]{Mohler:2013rwa}%
  \BibitemOpen
  \bibfield  {author} {\bibinfo {author} {\bibfnamefont {D.}~\bibnamefont {Mohler}}, \bibinfo {author} {\bibfnamefont {C.~B.}\ \bibnamefont {Lang}}, \bibinfo {author} {\bibfnamefont {L.}~\bibnamefont {Leskovec}}, \bibinfo {author} {\bibfnamefont {S.}~\bibnamefont {Prelovsek}},\ and\ \bibinfo {author} {\bibfnamefont {R.~M.}\ \bibnamefont {Woloshyn}},\ }\bibfield  {title} {\bibinfo {title} {{$D_{s0}^*(2317)$ Meson and $D$-Meson-Kaon Scattering from Lattice QCD}},\ }\href {https://doi.org/10.1103/PhysRevLett.111.222001} {\bibfield  {journal} {\bibinfo  {journal} {Phys. Rev. Lett.}\ }\textbf {\bibinfo {volume} {111}},\ \bibinfo {pages} {222001} (\bibinfo {year} {2013})},\ \Eprint {https://arxiv.org/abs/1308.3175} {arXiv:1308.3175 [hep-lat]} \BibitemShut {NoStop}%
\bibitem [{\citenamefont {Lang}\ \emph {et~al.}(2014)\citenamefont {Lang}, \citenamefont {Leskovec}, \citenamefont {Mohler}, \citenamefont {Prelovsek},\ and\ \citenamefont {Woloshyn}}]{Lang:2014yfa}%
  \BibitemOpen
  \bibfield  {author} {\bibinfo {author} {\bibfnamefont {C.~B.}\ \bibnamefont {Lang}}, \bibinfo {author} {\bibfnamefont {L.}~\bibnamefont {Leskovec}}, \bibinfo {author} {\bibfnamefont {D.}~\bibnamefont {Mohler}}, \bibinfo {author} {\bibfnamefont {S.}~\bibnamefont {Prelovsek}},\ and\ \bibinfo {author} {\bibfnamefont {R.~M.}\ \bibnamefont {Woloshyn}},\ }\bibfield  {title} {\bibinfo {title} {{$D_s$ Mesons with $DK$ and $D^*K$ Scattering Near Threshold}},\ }\href {https://doi.org/10.1103/PhysRevD.90.034510} {\bibfield  {journal} {\bibinfo  {journal} {Phys. Rev. D}\ }\textbf {\bibinfo {volume} {90}},\ \bibinfo {pages} {034510} (\bibinfo {year} {2014})},\ \Eprint {https://arxiv.org/abs/1403.8103} {arXiv:1403.8103 [hep-lat]} \BibitemShut {NoStop}%
\bibitem [{\citenamefont {Bali}\ \emph {et~al.}(2017)\citenamefont {Bali}, \citenamefont {Collins}, \citenamefont {Cox},\ and\ \citenamefont {Sch{\"a}fer}}]{Bali:2017pdv}%
  \BibitemOpen
  \bibfield  {author} {\bibinfo {author} {\bibfnamefont {G.~S.}\ \bibnamefont {Bali}}, \bibinfo {author} {\bibfnamefont {S.}~\bibnamefont {Collins}}, \bibinfo {author} {\bibfnamefont {A.}~\bibnamefont {Cox}},\ and\ \bibinfo {author} {\bibfnamefont {A.}~\bibnamefont {Sch{\"a}fer}},\ }\bibfield  {title} {\bibinfo {title} {{Masses and decay constants of the $D_{s0}^*(2317)$ and $D_{s1}(2460)$ from $N_f=2$ lattice QCD close to the physical point}},\ }\href {https://doi.org/10.1103/PhysRevD.96.074501} {\bibfield  {journal} {\bibinfo  {journal} {Phys. Rev. D}\ }\textbf {\bibinfo {volume} {96}},\ \bibinfo {pages} {074501} (\bibinfo {year} {2017})},\ \Eprint {https://arxiv.org/abs/1706.01247} {arXiv:1706.01247 [hep-lat]} \BibitemShut {NoStop}%
\bibitem [{\citenamefont {Cheung}\ \emph {et~al.}(2021)\citenamefont {Cheung}, \citenamefont {Thomas}, \citenamefont {Wilson}, \citenamefont {Moir}, \citenamefont {Peardon},\ and\ \citenamefont {Ryan}}]{Cheung:2020mql}%
  \BibitemOpen
  \bibfield  {author} {\bibinfo {author} {\bibfnamefont {G.~K.~C.}\ \bibnamefont {Cheung}}, \bibinfo {author} {\bibfnamefont {C.~E.}\ \bibnamefont {Thomas}}, \bibinfo {author} {\bibfnamefont {D.~J.}\ \bibnamefont {Wilson}}, \bibinfo {author} {\bibfnamefont {G.}~\bibnamefont {Moir}}, \bibinfo {author} {\bibfnamefont {M.}~\bibnamefont {Peardon}},\ and\ \bibinfo {author} {\bibfnamefont {S.~M.}\ \bibnamefont {Ryan}} (\bibinfo {collaboration} {Hadron Spectrum}),\ }\bibfield  {title} {\bibinfo {title} {{$DK \,I = 0$, $D\bar K\, I = 0, 1$ scattering and the $D_{s0}^*(2317)$ from lattice QCD}},\ }\href {https://doi.org/10.1007/JHEP02(2021)100} {\bibfield  {journal} {\bibinfo  {journal} {JHEP}\ }\textbf {\bibinfo {volume} {02}},\ \bibinfo {pages} {100}},\ \Eprint {https://arxiv.org/abs/2008.06432} {arXiv:2008.06432 [hep-lat]} \BibitemShut {NoStop}%
\bibitem [{\citenamefont {Mart{\'\i}nez~Torres}\ \emph {et~al.}(2015)\citenamefont {Mart{\'\i}nez~Torres}, \citenamefont {Oset}, \citenamefont {Prelovsek},\ and\ \citenamefont {Ramos}}]{MartinezTorres:2014kpc}%
  \BibitemOpen
  \bibfield  {author} {\bibinfo {author} {\bibfnamefont {A.}~\bibnamefont {Mart{\'\i}nez~Torres}}, \bibinfo {author} {\bibfnamefont {E.}~\bibnamefont {Oset}}, \bibinfo {author} {\bibfnamefont {S.}~\bibnamefont {Prelovsek}},\ and\ \bibinfo {author} {\bibfnamefont {A.}~\bibnamefont {Ramos}},\ }\bibfield  {title} {\bibinfo {title} {{Reanalysis of lattice QCD spectra leading to the $D_{s0}^*(2317)$ and $D_{s1}^*(2460)$}},\ }\href {https://doi.org/10.1007/JHEP05(2015)153} {\bibfield  {journal} {\bibinfo  {journal} {JHEP}\ }\textbf {\bibinfo {volume} {05}},\ \bibinfo {pages} {153}},\ \Eprint {https://arxiv.org/abs/1412.1706} {arXiv:1412.1706 [hep-lat]} \BibitemShut {NoStop}%
\bibitem [{\citenamefont {Gamermann}\ \emph {et~al.}(2010)\citenamefont {Gamermann}, \citenamefont {Nieves}, \citenamefont {Oset},\ and\ \citenamefont {Ruiz~Arriola}}]{Gamermann:2009uq}%
  \BibitemOpen
  \bibfield  {author} {\bibinfo {author} {\bibfnamefont {D.}~\bibnamefont {Gamermann}}, \bibinfo {author} {\bibfnamefont {J.}~\bibnamefont {Nieves}}, \bibinfo {author} {\bibfnamefont {E.}~\bibnamefont {Oset}},\ and\ \bibinfo {author} {\bibfnamefont {E.}~\bibnamefont {Ruiz~Arriola}},\ }\bibfield  {title} {\bibinfo {title} {{Couplings in coupled channels versus wave functions: application to the X(3872) resonance}},\ }\href {https://doi.org/10.1103/PhysRevD.81.014029} {\bibfield  {journal} {\bibinfo  {journal} {Phys. Rev. D}\ }\textbf {\bibinfo {volume} {81}},\ \bibinfo {pages} {014029} (\bibinfo {year} {2010})},\ \Eprint {https://arxiv.org/abs/0911.4407} {arXiv:0911.4407 [hep-ph]} \BibitemShut {NoStop}%
\bibitem [{\citenamefont {Oller}\ and\ \citenamefont {Meissner}(2001)}]{Oller:2000fj}%
  \BibitemOpen
  \bibfield  {author} {\bibinfo {author} {\bibfnamefont {J.~A.}\ \bibnamefont {Oller}}\ and\ \bibinfo {author} {\bibfnamefont {U.~G.}\ \bibnamefont {Meissner}},\ }\bibfield  {title} {\bibinfo {title} {{Chiral dynamics in the presence of bound states: Kaon nucleon interactions revisited}},\ }\href {https://doi.org/10.1016/S0370-2693(01)00078-8} {\bibfield  {journal} {\bibinfo  {journal} {Phys. Lett. B}\ }\textbf {\bibinfo {volume} {500}},\ \bibinfo {pages} {263} (\bibinfo {year} {2001})},\ \Eprint {https://arxiv.org/abs/hep-ph/0011146} {arXiv:hep-ph/0011146} \BibitemShut {NoStop}%
\bibitem [{\citenamefont {Carrasco}\ \emph {et~al.}(1993)\citenamefont {Carrasco}, \citenamefont {Nieves},\ and\ \citenamefont {Oset}}]{Carrasco:1991we}%
  \BibitemOpen
  \bibfield  {author} {\bibinfo {author} {\bibfnamefont {R.~C.}\ \bibnamefont {Carrasco}}, \bibinfo {author} {\bibfnamefont {J.}~\bibnamefont {Nieves}},\ and\ \bibinfo {author} {\bibfnamefont {E.}~\bibnamefont {Oset}},\ }\bibfield  {title} {\bibinfo {title} {{Coherent (gamma, pi0) photoproduction in a local approximation to the delta hole model}},\ }\href {https://doi.org/10.1016/0375-9474(93)90005-I} {\bibfield  {journal} {\bibinfo  {journal} {Nucl. Phys. A}\ }\textbf {\bibinfo {volume} {565}},\ \bibinfo {pages} {797} (\bibinfo {year} {1993})}\BibitemShut {NoStop}%
\bibitem [{\citenamefont {Boffi}\ \emph {et~al.}(1991)\citenamefont {Boffi}, \citenamefont {Bracci},\ and\ \citenamefont {Christillin}}]{Boffi:1991nh}%
  \BibitemOpen
  \bibfield  {author} {\bibinfo {author} {\bibfnamefont {S.}~\bibnamefont {Boffi}}, \bibinfo {author} {\bibfnamefont {L.}~\bibnamefont {Bracci}},\ and\ \bibinfo {author} {\bibfnamefont {P.}~\bibnamefont {Christillin}},\ }\bibfield  {title} {\bibinfo {title} {{Coherent (gamma, pi0) on nuclei}},\ }\href {https://doi.org/10.1007/BF02820558} {\bibfield  {journal} {\bibinfo  {journal} {Nuovo Cim. A}\ }\textbf {\bibinfo {volume} {104}},\ \bibinfo {pages} {843} (\bibinfo {year} {1991})}\BibitemShut {NoStop}%
\bibitem [{\citenamefont {Yamagata-Sekihara}\ \emph {et~al.}(2011)\citenamefont {Yamagata-Sekihara}, \citenamefont {Nieves},\ and\ \citenamefont {Oset}}]{Yamagata-Sekihara:2010kpd}%
  \BibitemOpen
  \bibfield  {author} {\bibinfo {author} {\bibfnamefont {J.}~\bibnamefont {Yamagata-Sekihara}}, \bibinfo {author} {\bibfnamefont {J.}~\bibnamefont {Nieves}},\ and\ \bibinfo {author} {\bibfnamefont {E.}~\bibnamefont {Oset}},\ }\bibfield  {title} {\bibinfo {title} {{Couplings in coupled channels versus wave functions in the case of resonances: application to the two $\Lambda(1405)$ states}},\ }\href {https://doi.org/10.1103/PhysRevD.83.014003} {\bibfield  {journal} {\bibinfo  {journal} {Phys. Rev. D}\ }\textbf {\bibinfo {volume} {83}},\ \bibinfo {pages} {014003} (\bibinfo {year} {2011})},\ \Eprint {https://arxiv.org/abs/1007.3923} {arXiv:1007.3923 [hep-ph]} \BibitemShut {NoStop}%
\bibitem [{\citenamefont {Montesinos}\ \emph {et~al.}(2024)\citenamefont {Montesinos}, \citenamefont {Song}, \citenamefont {Liang}, \citenamefont {Oset}, \citenamefont {Nieves},\ and\ \citenamefont {Albaladejo}}]{Montesinos:2024eoy}%
  \BibitemOpen
  \bibfield  {author} {\bibinfo {author} {\bibfnamefont {V.}~\bibnamefont {Montesinos}}, \bibinfo {author} {\bibfnamefont {J.}~\bibnamefont {Song}}, \bibinfo {author} {\bibfnamefont {W.-H.}\ \bibnamefont {Liang}}, \bibinfo {author} {\bibfnamefont {E.}~\bibnamefont {Oset}}, \bibinfo {author} {\bibfnamefont {J.}~\bibnamefont {Nieves}},\ and\ \bibinfo {author} {\bibfnamefont {M.}~\bibnamefont {Albaladejo}},\ }\bibfield  {title} {\bibinfo {title} {{Study of possible $DND^*$ bound states}},\ }\href {https://doi.org/10.1103/PhysRevD.110.054043} {\bibfield  {journal} {\bibinfo  {journal} {Phys. Rev. D}\ }\textbf {\bibinfo {volume} {110}},\ \bibinfo {pages} {054043} (\bibinfo {year} {2024})},\ \Eprint {https://arxiv.org/abs/2405.09467} {arXiv:2405.09467 [hep-ph]} \BibitemShut {NoStop}%
\bibitem [{\citenamefont {Liu}\ \emph {et~al.}(2023)\citenamefont {Liu}, \citenamefont {Lu},\ and\ \citenamefont {Geng}}]{Liu:2023uly}%
  \BibitemOpen
  \bibfield  {author} {\bibinfo {author} {\bibfnamefont {Z.-W.}\ \bibnamefont {Liu}}, \bibinfo {author} {\bibfnamefont {J.-X.}\ \bibnamefont {Lu}},\ and\ \bibinfo {author} {\bibfnamefont {L.-S.}\ \bibnamefont {Geng}},\ }\bibfield  {title} {\bibinfo {title} {{Study of the $DK$ interaction with femtoscopic correlation functions}},\ }\href {https://doi.org/10.1103/PhysRevD.107.074019} {\bibfield  {journal} {\bibinfo  {journal} {Phys. Rev. D}\ }\textbf {\bibinfo {volume} {107}},\ \bibinfo {pages} {074019} (\bibinfo {year} {2023})},\ \Eprint {https://arxiv.org/abs/2302.01046} {arXiv:2302.01046 [hep-ph]} \BibitemShut {NoStop}%
\bibitem [{\citenamefont {Liu}\ \emph {et~al.}(2025)\citenamefont {Liu}, \citenamefont {Pan}, \citenamefont {Liu}, \citenamefont {Wu}, \citenamefont {Lu},\ and\ \citenamefont {Geng}}]{Liu:2024uxn}%
  \BibitemOpen
  \bibfield  {author} {\bibinfo {author} {\bibfnamefont {M.-Z.}\ \bibnamefont {Liu}}, \bibinfo {author} {\bibfnamefont {Y.-W.}\ \bibnamefont {Pan}}, \bibinfo {author} {\bibfnamefont {Z.-W.}\ \bibnamefont {Liu}}, \bibinfo {author} {\bibfnamefont {T.-W.}\ \bibnamefont {Wu}}, \bibinfo {author} {\bibfnamefont {J.-X.}\ \bibnamefont {Lu}},\ and\ \bibinfo {author} {\bibfnamefont {L.-S.}\ \bibnamefont {Geng}},\ }\bibfield  {title} {\bibinfo {title} {{Three ways to decipher the nature of exotic hadrons: Multiplets, three-body hadronic molecules, and correlation functions}},\ }\href {https://doi.org/10.1016/j.physrep.2024.12.001} {\bibfield  {journal} {\bibinfo  {journal} {Phys. Rept.}\ }\textbf {\bibinfo {volume} {1108}},\ \bibinfo {pages} {1} (\bibinfo {year} {2025})},\ \Eprint {https://arxiv.org/abs/2404.06399} {arXiv:2404.06399 [hep-ph]} \BibitemShut {NoStop}%
\bibitem [{\citenamefont {Ikeno}\ \emph {et~al.}(2023)\citenamefont {Ikeno}, \citenamefont {Toledo},\ and\ \citenamefont {Oset}}]{Ikeno:2023ojl}%
  \BibitemOpen
  \bibfield  {author} {\bibinfo {author} {\bibfnamefont {N.}~\bibnamefont {Ikeno}}, \bibinfo {author} {\bibfnamefont {G.}~\bibnamefont {Toledo}},\ and\ \bibinfo {author} {\bibfnamefont {E.}~\bibnamefont {Oset}},\ }\bibfield  {title} {\bibinfo {title} {{Model independent analysis of femtoscopic correlation functions: An application to the $D_{s0}^*(2317)$}},\ }\href {https://doi.org/10.1016/j.physletb.2023.138281} {\bibfield  {journal} {\bibinfo  {journal} {Phys. Lett. B}\ }\textbf {\bibinfo {volume} {847}},\ \bibinfo {pages} {138281} (\bibinfo {year} {2023})},\ \Eprint {https://arxiv.org/abs/2305.16431} {arXiv:2305.16431 [hep-ph]} \BibitemShut {NoStop}%
\bibitem [{\citenamefont {Su}\ \emph {et~al.}(2025)\citenamefont {Su}, \citenamefont {Lyu}, \citenamefont {Liang},\ and\ \citenamefont {Oset}}]{Su:2025aiz}%
  \BibitemOpen
  \bibfield  {author} {\bibinfo {author} {\bibfnamefont {P.-S.}\ \bibnamefont {Su}}, \bibinfo {author} {\bibfnamefont {W.-T.}\ \bibnamefont {Lyu}}, \bibinfo {author} {\bibfnamefont {W.-H.}\ \bibnamefont {Liang}},\ and\ \bibinfo {author} {\bibfnamefont {E.}~\bibnamefont {Oset}},\ }\href@noop {} {\bibinfo {title} {{The $D_{s0}^*(2317)^+$ decay to $D_s^+\pi^0$ and $D_s^{*+}\gamma$}}} (\bibinfo {year} {2025}),\ \Eprint {https://arxiv.org/abs/2512.07421} {arXiv:2512.07421 [hep-ph]} \BibitemShut {NoStop}%
\bibitem [{\citenamefont {Bando}\ \emph {et~al.}(1985)\citenamefont {Bando}, \citenamefont {Kugo}, \citenamefont {Uehara}, \citenamefont {Yamawaki},\ and\ \citenamefont {Yanagida}}]{Bando:1984ej}%
  \BibitemOpen
  \bibfield  {author} {\bibinfo {author} {\bibfnamefont {M.}~\bibnamefont {Bando}}, \bibinfo {author} {\bibfnamefont {T.}~\bibnamefont {Kugo}}, \bibinfo {author} {\bibfnamefont {S.}~\bibnamefont {Uehara}}, \bibinfo {author} {\bibfnamefont {K.}~\bibnamefont {Yamawaki}},\ and\ \bibinfo {author} {\bibfnamefont {T.}~\bibnamefont {Yanagida}},\ }\bibfield  {title} {\bibinfo {title} {{Is $\rho$ Meson a Dynamical Gauge Boson of Hidden Local Symmetry?}},\ }\href {https://doi.org/10.1103/PhysRevLett.54.1215} {\bibfield  {journal} {\bibinfo  {journal} {Phys. Rev. Lett.}\ }\textbf {\bibinfo {volume} {54}},\ \bibinfo {pages} {1215} (\bibinfo {year} {1985})}\BibitemShut {NoStop}%
\bibitem [{\citenamefont {Bando}\ \emph {et~al.}(1988)\citenamefont {Bando}, \citenamefont {Kugo},\ and\ \citenamefont {Yamawaki}}]{Bando:1987br}%
  \BibitemOpen
  \bibfield  {author} {\bibinfo {author} {\bibfnamefont {M.}~\bibnamefont {Bando}}, \bibinfo {author} {\bibfnamefont {T.}~\bibnamefont {Kugo}},\ and\ \bibinfo {author} {\bibfnamefont {K.}~\bibnamefont {Yamawaki}},\ }\bibfield  {title} {\bibinfo {title} {{Nonlinear Realization and Hidden Local Symmetries}},\ }\href {https://doi.org/10.1016/0370-1573(88)90019-1} {\bibfield  {journal} {\bibinfo  {journal} {Phys. Rept.}\ }\textbf {\bibinfo {volume} {164}},\ \bibinfo {pages} {217} (\bibinfo {year} {1988})}\BibitemShut {NoStop}%
\bibitem [{\citenamefont {Mei{\ss}ner}(1988)}]{Meissner:1987ge}%
  \BibitemOpen
  \bibfield  {author} {\bibinfo {author} {\bibfnamefont {U.~G.}\ \bibnamefont {Mei{\ss}ner}},\ }\bibfield  {title} {\bibinfo {title} {{Low-Energy Hadron Physics from Effective Chiral Lagrangians with Vector Mesons}},\ }\href {https://doi.org/10.1016/0370-1573(88)90090-7} {\bibfield  {journal} {\bibinfo  {journal} {Phys. Rept.}\ }\textbf {\bibinfo {volume} {161}},\ \bibinfo {pages} {213} (\bibinfo {year} {1988})}\BibitemShut {NoStop}%
\bibitem [{\citenamefont {Nagahiro}\ \emph {et~al.}(2009)\citenamefont {Nagahiro}, \citenamefont {Roca}, \citenamefont {Hosaka},\ and\ \citenamefont {Oset}}]{Nagahiro:2008cv}%
  \BibitemOpen
  \bibfield  {author} {\bibinfo {author} {\bibfnamefont {H.}~\bibnamefont {Nagahiro}}, \bibinfo {author} {\bibfnamefont {L.}~\bibnamefont {Roca}}, \bibinfo {author} {\bibfnamefont {A.}~\bibnamefont {Hosaka}},\ and\ \bibinfo {author} {\bibfnamefont {E.}~\bibnamefont {Oset}},\ }\bibfield  {title} {\bibinfo {title} {{Hidden gauge formalism for the radiative decays of axial-vector mesons}},\ }\href {https://doi.org/10.1103/PhysRevD.79.014015} {\bibfield  {journal} {\bibinfo  {journal} {Phys. Rev. D}\ }\textbf {\bibinfo {volume} {79}},\ \bibinfo {pages} {014015} (\bibinfo {year} {2009})},\ \Eprint {https://arxiv.org/abs/0809.0943} {arXiv:0809.0943 [hep-ph]} \BibitemShut {NoStop}%
\end{thebibliography}%
\end{document}